\documentclass[letterpaper,12pt]{article}
\usepackage[latin1]{inputenc} 
\usepackage{multirow}
\usepackage{enumitem}
\usepackage{bigstrut,rotating}
\usepackage[top=1in, bottom=1in, left=1in, right=1in]{geometry}
\usepackage{graphicx,rotating,amsthm,amsbsy,amssymb,amsmath, caption, natbib, subcaption, bbm, color, thmtools}
\usepackage[draft]{hyperref}

\usepackage{bm}
\usepackage{algorithm}
\usepackage[noend]{algpseudocode}
\usepackage{natbib}
\usepackage{makecell}
\usepackage{float}  
\usepackage{lipsum}
\usepackage{setspace}


\definecolor{darkgreen}{RGB}{0,100,0}
\definecolor{teal}{RGB}{0,128,128}

\newtheorem{theorem}{Theorem}
\newtheorem{definition}{Definition}
\newtheorem{proposition}{Proposition}

\usepackage{xr}

\makeatletter
\newcommand*{\addFileDependency}[1]{
  \typeout{(#1)}
  \@addtofilelist{#1}
  \IfFileExists{#1}{}{\typeout{No file #1.}}
}
\makeatother

\newcommand*{\myexternaldocument}[1]{%
    \externaldocument{#1}%
    \addFileDependency{#1.tex}%
    \addFileDependency{#1.aux}%
}
\myexternaldocument{SupplementaryMaterial_round2}

\title{Differentially Private Methods for Compositional Data}

\begin{document}

\begin{center}
    {\Large \textbf{Differentially Private Inference for Compositional Data}}  \\
    Qi Guo, Andr\'es F. Barrientos, and V\'ictor Pe\~na\footnote{Qi Guo graduated from the Department of Statistics, Florida State University, USA (qg17@fsu.edu);   Andr\'es F. Barrientos is Assistant Professor, Department of Statistics, Florida State University, USA (abarrientos@fsu.edu); V\'ictor Pe\~na  is Assistant Professor, Department d'Estad\'istica i Investigaci\'o Operativa,
    Universitat Polit\`ecnica de Catalunya
    Barcelona, Spain (victor.pena.pizarro@upc.edu).} \\
\end{center}

\begin{abstract}
Confidential data, such as electronic health records, activity data from wearable devices, and geolocation data, are becoming increasingly prevalent. Differential privacy provides a framework to conduct statistical analyses while mitigating the risk of leaking private information. Compositional data, which consist of vectors with positive components that add up to a constant, have received little attention in the differential privacy literature. This article proposes differentially private approaches for analyzing compositional data using the Dirichlet distribution. We explore several methods, including Bayesian and bootstrap procedures. For the Bayesian methods, we consider posterior inference techniques based on Markov Chain Monte Carlo, Approximate Bayesian Computation, and asymptotic approximations. We conduct an extensive simulation study to compare these approaches and make evidence-based recommendations. Finally, we apply the methodology to a data set from the American Time Use Survey. 

\end{abstract}

\section{Introduction}


A significant challenge in the statistical analysis of confidential data is the trade-off between obtaining accurate statistics and protecting sensitive information. Differential privacy (DP), proposed by \cite{dwork2006calibrating}, offers a formal framework for statistical analyses of confidential data that controls the risk of leaking private information.  

DP is a well-defined mathematical property of randomized algorithms. The outputs of DP methods are typically noisy versions of summary statistics that are robust against changes in individual data entries, minimizing the information attackers can learn about specific individuals.

This article proposes DP algorithms for compositional data -- vectors with positive components that sum up to a constant, typically one. Compositional data sets appear across many different disciplines: sociologists measure time spent on daily activities, chemists study chemical compositions in samples, and environmental scientists analyze material compositions of solid waste. For this reason, considerable attention has been devoted to developing methodologies for compositional data \citep{aitchison1982statistical,bacon2011short,ongaro2013generalization}.  

We propose and evaluate DP approaches for analyzing compositional data using the Dirichlet distribution as the statistical model. The Dirichlet distribution is convenient for analyzing compositional data due to its mathematical properties and ease of interpretation. Our methodology defines a DP summary statistic by adding random noise to a left-censored version of the sufficient statistic of the Dirichlet distribution.

\subsection{Related Work}

In recent years, several authors have developed statistical methods that are valid under DP constraints. Recent advancements include hypothesis testing for binomial data \citep{Awan_Slavkovic_2020}, inference for linear regression models \citep{barrientos2019differentially,pena2021differentially,ferrando2022parametric}, confidence intervals for the mean in normal models \citep{karwa2017finite}, and noise-aware Bayesian inference for linear regression \citep{bernstein2019differentially} and generalized linear models \citep{kulkarni2021differentially}. However, few existing methods can be applied to compositional data, which is the primary focus of this work. The technique proposed by \cite{bernstein2018differentially}, which is applicable to models within the exponential family, can be used to analyze compositional data. Unfortunately, it requires computing integrals that are not analytically available and are computationally expensive to evaluate numerically. Another related article is \cite{ferrando2022parametric}, which uses parametric bootstrap to produce DP confidence intervals for distributions within the exponential family. Their work assumes that the support of the sufficient statistic is bounded, but this condition is not satisfied for the Dirichlet distribution.

\subsection{Main contributions} 

Our primary contributions are as follows:
\begin{itemize}
    \item We propose Bayesian and frequentist methods for analyzing compositional data under DP constraints. The methods are based on the Dirichlet distribution. DP is achieved through censoring and perturbing sufficient statistics.
    \item We propose a DP approach to select the censoring threshold. This approach is guaranteed to select the threshold we would choose without DP constraints as the sample size grows.
    \item We describe how to set prior distributions appropriately. This is important because vague priors tend to perform poorly in this context.
    \item We describe and compare algorithms for implementing the Bayesian methods. We consider algorithms based on Markov Chain Monte Carlo, scalable strategies based on data-splitting, Approximate Bayesian Computation algorithms, and an asymptotic approximation proposed in \cite{bernstein2018differentially}. 
    \item We provide recommendations for implementing these methods, taking into account modeling preferences and computational resources.
\end{itemize}

\section{Background}

This section introduces the formal definition of DP and highlights key properties that underpin our methodology. Then, we provide an introduction to the Dirichlet distribution and how to infer its unknown parameter.


\subsection{Differential privacy}\label{subsec:backgroundDP}

To define DP formally, we first introduce the concept of neighboring data sets. Data sets $\boldsymbol{D}$ and $\boldsymbol{D}'$ are considered neighbors if they are of the same size and differ in only one observation.

DP ensures that outputs for neighboring data sets are similar, making it difficult for attackers to distinguish whether a given output was computed based on $\boldsymbol{D}$ or $\boldsymbol{D}'$. Let $S$ be a random mechanism that takes as input a data set $\boldsymbol{D} = \{ \boldsymbol{x}_i\}_{i=1}^n$, where $\boldsymbol{x}_i$ represents the $i$th individual in the sample. The similarity between $S(\boldsymbol{D})$ and $S(\boldsymbol{D}')$ ensures that minimal information can be learned about the difference between $\boldsymbol{D}$ and $\boldsymbol{D}'$. In DP, this similarity is controlled by a parameter $\epsilon$, known as the privacy budget. The privacy budget $\epsilon$ controls the degree of privacy offered by $S$, with lower values of $\epsilon$ implying higher privacy levels. Now, we can proceed to the formal definition of DP.

\begin{definition}{\bf{Differential Privacy.}}\label{DPdef} Given $\epsilon > 0$, a random mechanism $S$ is $\epsilon$-DP if for all pairs of neighboring data sets $(\boldsymbol{D},\boldsymbol{D}')$, and for every $A \subseteq {\rm Range}(S)$,
$
    {\rm Pr}\left[S(\boldsymbol{D})\in A | \boldsymbol{D} \right]\leq \exp(\epsilon){\rm Pr}\left[S(\boldsymbol{D}') \in A | \boldsymbol{D}' \right].
$
\end{definition}
In Definition~\ref{DPdef}, the data sets $(\boldsymbol{D},\boldsymbol{D}')$ are treated as non-random objects. As $\epsilon$ decreases, the probability distributions of $S(\boldsymbol{D})$ and $S(\boldsymbol{D}')$ become increasingly similar (i.e., the privacy level increases).

DP has several properties that make it particularly useful when designing statistical methods. Three relevant properties are post-processing, sequential composition, and parallel composition.

\begin{proposition}\label{prop:postprocessing}{\bf{Post-processing.}} Given $S$ that satisfies $\epsilon$-DP and for any function $T$ defined on ${\rm Range}(S)$, the composition $T(S(\cdot))$ satisfies $\epsilon$-DP.
\end{proposition}
\begin{proposition}\label{prop:seqcomposition}{\bf{Sequential composition.}} Let $S_1$ and $S_2$ be $\epsilon_1$- and $\epsilon_2$-DP mechanisms, respectively. Then, the random mechanism $(S_1(\boldsymbol{D}), \, S_2(S_1(\boldsymbol{D}),\boldsymbol{D}))$
satisfies $(\epsilon_1+\epsilon_2)$-DP.
\end{proposition}
\begin{proposition}\label{prop:parcomposition}{\bf{Parallel composition.}} Let $S_1,\, ... \,,S_K$ be $K$ mechanisms that satisfy $\epsilon_1$-DP, $\cdots$, $\epsilon_K$-DP, respectively. Then, the joint mechanism  $(S_1(\boldsymbol{D}_1), \, ... \,, \, S_K(\boldsymbol{D}_K))$, where $\boldsymbol{D}_1 \cap\cdots\cap \boldsymbol{D}_K=\varnothing$ and $\boldsymbol{D}_1 \cup \boldsymbol{D}_2 \cup\, ... \,\cup \boldsymbol{D}_k \subset \boldsymbol{D}$ for $k \in \{ 1, \, ... \, ,K\}$,
satisfies $\max_{k\in \{1,\cdots,K\}} \epsilon_k$-DP.
\end{proposition}

Proposition \ref{prop:postprocessing} implies that transforming the output of an $\epsilon$-DP mechanism does not incur on extra loss of privacy. Sequential composition (see Proposition \ref{prop:seqcomposition}) is a key property that quantifies the total privacy cost when multiple queries on $\boldsymbol{D}$ are requested. Parallel composition, as described in Proposition \ref{prop:parcomposition}, enables the modular design of mechanisms: if all the components of a mechanism are differentially private on disjoint data sets, then so is their composition, and the total privacy cost is upper bounded by $\max_{k\in \{1,2,\cdots,K\}} \epsilon_k$. 

Two privacy-ensuring mechanisms are relevant for this work: the Laplace and the Geometric mechanisms. To define them, we assume that our goal is to design a DP version of a confidential summary statistic denoted by $S_0$. To produce this DP statistic, we must compute the global sensitivity of $S_0$, which is an upper bound on the maximum change (over all possible $ \boldsymbol{D}$) that $S_0(\boldsymbol{D})$ can experience when a single observation is added to or removed from $\boldsymbol{D}$. 

\begin{definition}\label{def:GlobalSensitivity} {\bf{Global Sensitivity.}}
The global sensitivity of the summary $S_0$, denoted by ${\rm GS}(S_0)$, is defined as
    ${\rm GS}(S_0) = \sup_{(\boldsymbol{D},\boldsymbol{D}')} \lVert S_0(\boldsymbol{D})-S_0(\boldsymbol{D}')\rVert_1$,
where $\boldsymbol{D}$ and $\boldsymbol{D}'$ are neighboring data sets.
\end{definition}

The \emph{Laplace mechanism}, proposed in \cite{dwork2006calibrating}, defines an $\epsilon$-DP version of $S_0$ by adding a Laplace-distributed perturbation term to $S_0(\boldsymbol{D})$. Specifically, for a real-valued function $S_0:\boldsymbol{D} \to \mathbb{R}^d$ with global sensitivity ${\rm GS}(S_0)$ and privacy budget $\epsilon$, the output of the mechanism is 
    $S_{L}(\boldsymbol{D}) := S_0(\boldsymbol{D}) + \boldsymbol{\varepsilon}^{L}$,
where $\boldsymbol{\varepsilon}^{L}$ is a $d$-dimensional vector with entries independently sampled from ${\rm Laplace}(0,{\rm GS}(S_0)/\epsilon)$. The ${\rm Laplace}(m,s)$ distribution is parameterized by $(m,s) \in \mathbb{R} \times \mathbb{R}^+ $ and its probability density function is
     $p_{\rm Lap}(\varepsilon^{L} \, | \, m,s) = \exp\left(-{|\varepsilon^{L}-m|}/{s}\right)/{2s},$  for $\varepsilon^L \in \mathbb{R}$.

The \emph{Geometric mechanism}, proposed by \cite{ghosh2012universally}, is a discretized version of the Laplace mechanism. The Geometric mechanism adds random noise drawn from the two-sided geometric distribution, also known as the discrete Laplace distribution \citep{inusah2006discrete}, to a discrete summary statistic. Specifically, for an integer-valued function $S_0:\boldsymbol{D} \to \mathbb{Z}^d$ with global sensitivity $\mathrm{GS}(S_0)$ and privacy budget $\epsilon$, the mechanism outputs 
    $S_{G}(\boldsymbol{D}) := S_0(\boldsymbol{D}) + \boldsymbol{\varepsilon}^{G}$,
where $\boldsymbol{\varepsilon}^{G}$ is a $d$-dimensional vector with entries independently sampled from ${\rm TwoSidedGeometric}(\exp(-\epsilon /{\rm GS}(S_0)))$. The probability mass function of the ${\rm TwoSidedGeometric}(t)$ distribution with parameter $t \in (0,1)$ is
    $p_{\rm TwoGeo}(\varepsilon^{G} \, | \, t) = t^{|\varepsilon^{G}|} \,  (1-t)/(1+t) ,$ for $\varepsilon^{G} \in \mathbb{Z}$.

\subsection{Compositional data and the Dirichlet distribution}\label{subsec:compositionaldata}

In this section, we review well-known facts about the Dirichlet distribution that are useful for our purposes. We begin by setting the notation. Let us assume that we have information on $n$ individuals, where each observation is compositional. For each individual $i \in \{ 1,\dots,n \}$, the corresponding compositional observation $\boldsymbol{x}_i=(x_{i1},\dots,x_{id})^T$ is a vector taking values on the $(d-1)$-dimensional simplex  
$
\Delta_{d-1} = \left\{(x_1, \, ... \, , x_d)^T : x_j \geq 0, \sum_{j=1}^d x_j=1\right\}.
$

To model $\boldsymbol{D} = \{ \boldsymbol{x}_i\}_{i=1}^n$, we assume that $\boldsymbol{x}_1,\dots,\boldsymbol{x}_n$ are independent and identically distributed samples from a Dirichlet distribution with parameter $\boldsymbol{\alpha} = (\alpha_1,\dots,\alpha_d)^T$, which we denote by
$\boldsymbol{x}_i \, | \, \boldsymbol{\alpha}\overset{\rm iid}{\sim}{\rm Dirichlet}( \boldsymbol{\alpha}).$
It is straightforward to see that the Dirichlet distribution is a member of the exponential family and that
$S_0(\boldsymbol{D}) = n^{-1} \left(\sum_{i=1}^n \log x_{i1}, \, ... \, ,\sum_{i=1}^n \log x_{id} \right)^T$ is a sufficient statistic.

We consider Bayesian and frequentist approaches to make inferences about $\boldsymbol{\alpha}$. In both paradigms, the sufficient statistic $S_0(\boldsymbol{D})$ is enough to perform full inference and prediction. 

For frequentist inference, we focus on the maximum likelihood estimator (MLE) of $\boldsymbol{\alpha}$. Given that the MLE cannot be computed analytically, \cite{minka2000estimating} proposed a convergent fixed-point iteration algorithm for estimating $\boldsymbol{\alpha}$. 

Bayesian inference requires specifying a prior probability distribution on $\boldsymbol{\alpha}$ representing the available prior information about this parameter. Since $\boldsymbol{\alpha}$ takes values on $\mathbb{R}^d_+$,  the prior distribution, denoted as $\pi_0$, is also defined on $\mathbb{R}^d_+$. The posterior distribution of $\boldsymbol{\alpha}$ is given by
\begin{equation}\label{eq:BayesModel}
        \pi(\boldsymbol{\alpha} \, | \, S_0(\boldsymbol{D}))  \propto  e^{(\boldsymbol{\alpha}-\boldsymbol{1})^T n S_0(\boldsymbol{D}) -C(\boldsymbol{\alpha})}  \pi_0(\boldsymbol{\alpha}), \, \, \, \boldsymbol{\alpha} \in \mathbb{R}^d_+,
\end{equation}
where $\boldsymbol{1}$ is a column vector with $d$ ones and $C(\boldsymbol{\alpha})$ is the cumulant-generating function of the distribution.

\section{Ensuring differential privacy}

This section outlines the process of ensuring differential privacy through left-censoring the sufficient statistic and adding a perturbation term. We then present a DP strategy for selecting the censoring threshold, followed by a description of the algorithm used to release the DP statistic.

{\color{black} The supplementary material (Section~\ref{sec:supp_cens}) contains an analysis of how censoring the sufficient statistic impacts our inferences. Our findings indicate that inferences based on censored data closely align with uncensored results when the threshold is small and $\sum_{j = 1}^d \alpha_j$ is not near zero.}

\subsection{Differentially private sufficient statistic}\label{sec:DPsufficient}

Creating a DP version of $S_0(\boldsymbol{D})$ cannot be achieved by directly applying the Laplace mechanism due to its unbounded support. The entries of any compositional datum can be arbitrarily close to zero, causing the logarithm to potentially diverge to $-\infty$.

To bound the global sensitivity of $S_0$, we propose left-censoring the observations at a small value $a \in (0,1)$, defining $\tilde{\boldsymbol{x}}_i = (\max\{x_{i1},a\},\, ... \, ,\max\{x_{id},a\})$ and $\tilde{\boldsymbol{D}} = \{ \tilde{\boldsymbol{x}}_i\}_{i=1}^n$. After applying the left-censoring to the observations, the global sensitivity becomes 
{\color{black} \begin{align*}
{\rm GS}(S_0) & =  \sup_{(\tilde{\boldsymbol{D}},\tilde{\boldsymbol{D}}') \, \, \text{neigh.}} \lVert S_0(\tilde{\boldsymbol{D}})-S_0(\tilde{\boldsymbol{D}}')\rVert_1 \\
& =  
n^{-1}\sup_{\boldsymbol{x},\boldsymbol{x}' \in \Delta_{d-1}} \sum_{j=1}^d|\log(\max\{x_{j},a\})-\log(\max\{x_{j}',a\})| \\
 & \leq  - n^{-1} d \log(a),
 \end{align*}
where $\tilde{\boldsymbol{D}}$ and $\tilde{\boldsymbol{D}}'$ are neighboring data sets.} To make the dependence on $a$ explicit, we denote the censored sufficient statistic as $S_0(\boldsymbol{D}, a) := S_0(\tilde{\boldsymbol{D}})$. 

After defining $S_0(\boldsymbol{D}, a)$, we can apply the Laplace mechanism. We define $S_{L}(\boldsymbol{D},a) = S_0(\boldsymbol{D},a) + \boldsymbol{\varepsilon}^{L}$, where the entries of $\boldsymbol{\varepsilon}^{L}$ are  sampled from 
${\rm Laplace}(0, -d \log a/(n \epsilon_1))$. The statistic $S_{L}(\boldsymbol{D},a)$ is an $\epsilon_1$-DP version of $S_0(\boldsymbol{D},a)$, and our goal is to make inferences about $\boldsymbol{\alpha}$ based on  $S_{L}(\boldsymbol{D},a)$. Clearly, $S_{L}(\boldsymbol{D},a)$ converges in probability to $S_0(\boldsymbol{D},a)$ as the sample size $n$ increases for a fixed privacy budget $\epsilon_1$.

\subsection{Selecting the threshold for censoring the sufficient statistic}\label{sec:DP_thershold}

Since $S_{L}(\boldsymbol{D},a)$ depends on $a$, we develop an algorithm to select it. There is a bias-variance trade-off in selecting $a$.  Small values of $a$ imply $S_0(\boldsymbol{D},a) \approx S_0(\boldsymbol{D})$, but they lead to large variances in $\boldsymbol{\varepsilon}^{L}$. Conversely, large values of $a$ induce substantial bias, but lead to small variances in $\boldsymbol{\varepsilon}^{L}$. 

{\color{black} We propose to select $a$ from a list of $M$ candidates. The procedure begins with the analyst specifying a desired proportion $t_c$ of observations $\boldsymbol{x}_i \in \mathbb{R}^d$ such that at least one of the $d$ components is censored. Telling the user that $t_c$ is achieved exactly could leak private information. For that reason, we develop a DP algorithm that gives noisy estimates of the proportion of censored observations and can identify if any of the $M$ candidates for $a$ is likely to achieve censoring rates that are similar or lower than $t_c$.

For a given user-specified rate $t_c$ and candidates $a_1, \, ... \, , a_M$ such that $0 = a_0 < a_1 < ... < a_M < a_{M+1} = 1$, let $p_m$ be the probability that $\boldsymbol{x} = (x_1, ..., x_d) \sim {\rm Dirichlet}(\boldsymbol{\alpha})$ is subject to censoring if $a_m$ is used. In other words, $p_m = P(x_j < a_m$ for some $j \in {1, ..., d})$. We define the optimum $a_{\rm opt}$ to be the largest candidate that achieves the desired censoring rate, meaning $a_{\rm opt} = a_m$ if $p_m < t_c < p_{m+1}$. 

Let $s_m$ be the number of observations that would have censored components with $a_m$ but not with $a_{m-1}$, namely $
s_m = \sum_{i=1}^n \mathbb{I}(a_{m-1} \leq \min_j{x_{i,j}} < a_{m})$ for $m \in \{1, ..., M+1\}$.

If we didn't have privacy constraints, we could find $a_{\rm opt}$ with the proportion $\hat p_m = n^{-1}\sum_{l=1}^m s_l$. However, under privacy constraints, we need to produce a noisy DP version of $p_m$. Given that $(s_1, ..., s_{M+1})$ is a vector with discrete components and global sensitivity equal to 2, we can use the Geometric mechanism to define 
$
S_{G}(\boldsymbol{D}) = (s_{G,1},\, ... \,, s_{G,M+1}) = (\max\{0, s_1 +\varepsilon_1^G\} , \, ... \,, \max\{0, s_{M+1}+\varepsilon_{M+1}^G\}),
$
where $\varepsilon_1^G, \, ... \, , \varepsilon_{m+1}^G$ are independently sampled from ${\rm TwoSidedGeometric}(\exp(-\epsilon_2/2))$. 

We can estimate $a_{\text{opt}}$ with $
\hat p_{G,m} = \sum_{l=1}^m s_{G,l}/\sum_{l=1}^{M+1} s_{G,l}.$ Without privacy constraints, we would select $a_{\mathrm{opt}}$  as the $a_l$ such that 
 $p_l \leq t_c < p_{l+1}$. If there is no $a_l$ for $l \ge 1$ that satisfies the condition, then $a_{\mathrm{opt}}= 0$. Under privacy constraints, we propose substituting the $p_l$ by their DP estimates $\hat{p}_{G,l}$, defining
$a(\boldsymbol{x}_1, ..., \boldsymbol{x}_n, \boldsymbol{\varepsilon}^G)$ as the $a_l$ 
such that $\hat p_{G,l} \leq t_c < \hat p_{G, l+1}$,
where $\boldsymbol{\varepsilon}^G = (\varepsilon_1^G, ..., \varepsilon_{M+1}^G)$. When none of the DP censoring rates $\hat p_{G,m}$ for the $M$ candidates is below the desired threshold $t_c$, then $a(\boldsymbol{x}_1, ..., \boldsymbol{x}_n, \boldsymbol{\varepsilon}^G)$ is equal to 0. In such a case, the user can assess whether to use $a_1$ depending on whether they are willing to accept $\hat p_{G,1}$ as the censoring rate.

As the sample size increases,  $a(\boldsymbol{x}_1, ..., \boldsymbol{x}_n, \boldsymbol{\varepsilon}^G)$ converges to $a_{\rm opt}$. 
The following theorem characterizes this convergence, including the order of the privacy budget $\epsilon_2$ required as a function of $n$ to achieve it. The proof of this theorem is provided in the supplementary material.

\begin{theorem}\label{theo:opt_a}
Assume $\epsilon_2 = \Omega(n^{-\gamma})$, with $\gamma \in [0,1)$, that is, there exists a constant \( C > 0 \) such that \( \epsilon_2 \geq C n^{-\gamma} \) for sufficiently large \( n \).
Then
$
a \equiv a(\boldsymbol{x}_1, ..., \boldsymbol{x}_n, \boldsymbol{\varepsilon}^G) 
$
converges in probability to $a_{\rm opt}$ as $n$ goes to $\infty$.
\end{theorem}

{\bf Remark}: The assumption of the data-generating mechanism being a Dirichlet distribution is not necessary for Theorem \ref{theo:opt_a} to hold. In fact, the theorem remains valid for any distribution on the simplex that has full support.}

\subsection{Releasing the differentially private statistic}

{\color{black} Algorithm \ref{alg:DPsummaries} describes the steps to follow to release $S_{G}(\boldsymbol{D})$, $(\hat p_{G,1}, ..., \hat p_{G,M})$, the threshold $a$,  and $S_{L}(\boldsymbol{D},a)$. The algorithm sets $a$ to be at least $a_1$}. For the Bayesian methods (in Section~\ref{sec:bayesian}), we argue that it can be convenient to split the data into subsets $\boldsymbol{D}_1$ and $\boldsymbol{D}_2$ to alleviate the effects of prior choice on posterior inference. Algorithm~\ref{alg:DPsummaries} considers this possibility.

We conclude this section with a theorem that states that Algorithm \ref{alg:DPsummaries} is $(\epsilon_1+\epsilon_2)$-DP. Its proof can be found in the Supplementary Material. 

\begin{theorem}\label{theo1} Algorithm \ref{alg:DPsummaries} satisfies $(\epsilon_1+\epsilon_2)$-DP. 
\end{theorem}

\begin{algorithm}
  \caption{$(\epsilon_1+\epsilon_2)$-DP algorithm to select $a$ and release DP statistic}\label{alg:DPsummaries}
  \label{alg:truncation}
  \begin{algorithmic}
  \Procedure{DPss}{$\boldsymbol{D},\epsilon_1,\epsilon_2,(a_1,a_2, \, ... \, ,a_M)$, {\color{black} $t_c$}, $n_1$, $n_2$}
    \State {\color{black} compute the counts $(s_1, ..., s_{M+1})$ using $\boldsymbol{D}$, $(a_1,a_2, \, ... \, ,a_M)$, and $t_c$}
    \State draw $\varepsilon^G_m \overset{\mathrm{iid}}{\sim} {\rm TwoSidedGeometric}(\exp(-\epsilon_2/2))$, $m=1,\, ... \, ,M+1$
    \State {\color{black} compute 
    \begin{eqnarray*}
        (s_{G,1},\, ... \,, s_{G,M+1}) &=& (\max\{0, s_1 +\varepsilon_1^G\} , ..., \max\{s_{M+1}+\varepsilon_{M+1}^G\}) \\
        \hat p_{G,m} &=& \left(\sum_{l=1}^{M+1} \min\{s_{G,l},0\}\right)^{-1}\left(\sum_{l=1}^m \min\{s_{G,l},0\}\right)   
    \end{eqnarray*}}
    \State {\color{black} define $
    a = 
    \begin{cases}
a_l & \text{if } \hat p_{G,l} \leq t_c < \hat p_{G, l+1}, \, l = 1, ..., M \\
a_1 & \text{otherwise }
\end{cases}
    $}
    \If{$n_1=0$} 
        \State draw $\varepsilon^L_j \overset{\mathrm{iid}}{\sim} {\rm Laplace}(0, - d \log(a)/(n \epsilon_1))$, $j \in \{1, \, ... \, ,d\}$
        \State compute $S_{L}(\boldsymbol{D},a) = S_0(\boldsymbol{D},a) + (\varepsilon^{L}_1,\dots,\varepsilon^{L}_d)$
        \State {\bf return} {\color{black} $S_{G}(\boldsymbol{D}) = (s_{G,1},\, ... \,, s_{G,M+1})$, $(\hat p_{G,1}, ..., \hat p_{G,M})$,} $a$, and $S_{L}(\boldsymbol{D},a)$
    \Else
    \State Split $\boldsymbol{D}$ uniformly at random into $\boldsymbol{D}_1$ {\color{black} of size $n_1$} and $\boldsymbol{D}_2$ {\color{black} of size $n_2$}
    \State draw $\varepsilon^L_{j,l} \overset{\mathrm{iid}}{\sim} {\rm Laplace}(0, -d \log(a)/(n_l \epsilon_1))$, $j \in \{1, \, ... \, ,d\}$,  $l \in \{1,2\}$
    \State compute $S_{L}(\boldsymbol{D}_l,a) = S_0(\boldsymbol{D}_l,a) + (\varepsilon^{L}_{1,l},\dots,\varepsilon^{L}_{d,l})$,  $l \in \{1,2\}$
    \State {\bf return} {\color{black} $S_{G}(\boldsymbol{D}) = (s_{G,1},\, ... \,, s_{G,M+1})$, $(\hat p_{G,1}, ..., \hat p_{G,M})$,} $a$, $S_{L}(\boldsymbol{D}_1,a)$, and $S_{L}(\boldsymbol{D}_2,a)$
    \EndIf 
    \EndProcedure
  \end{algorithmic}
\end{algorithm}

\section{Methods}

This section describes our proposed methods to make inferences about $\boldsymbol{\alpha}$ with DP constraints. We first introduce a frequentist bootstrap method and then we proceed to explain the Bayesian methodology. For the Bayesian methods, we consider a variety of algorithms for posterior inference.

\subsection{Frequentist methods}\label{sec:DPfreq}

As discussed in Section \ref{subsec:compositionaldata}, we can obtain the MLE for $\boldsymbol{\alpha}$ using the sufficient statistic $S_0(\boldsymbol{D})=S_0(\boldsymbol{D},0)$ and the convergent fixed-point iteration technique proposed by \cite{minka2000estimating}. Let ${\rm MLE}(S_0(\boldsymbol{D},0))$ be the function returning the MLE of $\boldsymbol{\alpha}$. Since $S_L(\boldsymbol{D},a)$ is an approximation of  $S_0(\boldsymbol{D},0)$, we could obtain a DP estimate of $\boldsymbol{\alpha}$ using ${\rm MLE}(S_L(\boldsymbol{D},a))$. However, to obtain proper inferences, we cannot omit the censoring and noise added when computing $S_L(\boldsymbol{D},a)$. To account for these aspects, we use the parametric bootstrap \citep{efron2012bayesian} to approximate the distribution of ${\rm MLE}(S_L(\boldsymbol{D},a))$.

To implement the parametric bootstrap, we first account for the noise added to $S_0(\boldsymbol{D},a)$. \textcolor{black}{Our proposal is subtracting off a random term $\boldsymbol{\varepsilon}^{*L}$, which is distributed similarly to the noise $\boldsymbol{\varepsilon}^{L}$ that was added to achieve DP, defining  $\widetilde{S_0}(\boldsymbol{D},a) = S_{L}(\boldsymbol{D},a) - \boldsymbol{\varepsilon}^{\ast L}$, while ensuring that $\widetilde{S_0}(\boldsymbol{D},a)$ resides within an appropriate space.} Then, we compute $\boldsymbol{\alpha}^* = {\rm MLE}(\widetilde{S_0}(\boldsymbol{D},a))$ and, to account for sampling error, generate a simulated data set $\widetilde{\boldsymbol{D}}$ of size $n$ using ${\rm Dirichlet}( \boldsymbol{\alpha}^*)$. Finally, we compute $S_0(\widetilde{\boldsymbol{D}},a)$, which accounts for the censoring, and obtain $\widetilde{\boldsymbol{\alpha}} = {\rm MLE}(S_0(\widetilde{\boldsymbol{D}},a))$. We use the distribution of $\widetilde{\boldsymbol{\alpha}}$ to approximate the sampling distribution of ${\rm MLE}(S_L(\boldsymbol{D},a))$. Algorithm \ref{alg:MLE result}, henceforth DPBoots, summarizes our strategy. 

Now, we justify why we propose Algorithm~\ref{alg:MLE result} as a parametric bootstrap algorithm. In the usual, non-private parametric bootstrap, we would sample from $
p_{\rm Dir}(\boldsymbol{x} \mid \boldsymbol{\alpha} = {\rm MLE}(S_0(\boldsymbol{D})))$. If we observe $S_L(\boldsymbol{D}) = S_0(\boldsymbol{D}) + \boldsymbol{\varepsilon}^{L}$ instead, we can rewrite the parametric bootstrap conditional on $\boldsymbol{\varepsilon}^{L}$ as
$
p_{\rm Dir}(\boldsymbol{x} \mid \boldsymbol{\alpha} = {\rm MLE}(S_L(\boldsymbol{D})-\boldsymbol{\varepsilon}^{L})),
$
where $S_L(\boldsymbol{D})-\boldsymbol{\varepsilon}^{L}$ must be in ${\rm Range}(S_0)$. {\color{black} Since $\exp(S_0) = \left(\prod_{i=1}^n x_{i1}^{1/n}, ..., \prod_{i=1}^n x_{id}^{1/n}\right)$  and $\sum_{j=1}^d \prod_{i=1}^n x_{ij}^{1/n} \leq \sum_{j=1}^d x_{i'j}^{1/n} \leq 1 $ for all $i' = 1,...,n$, then ${\rm Range}(S_0) = \{(s_{0,1},...,s_{0,d}) \, : \, \sum_{j}^d \exp(s_{0,d}) \leq 1\}$.
} Here, we assume that $S_0(\boldsymbol{D}) \approx S_0(\boldsymbol{D}, a)$, which is something we can verify using the DP score function $S_{G}(\boldsymbol{D})$. Evidence against such an assumption arises when the selected threshold is $a = a_1$ (the smallest one among the $M$ candidates), and  {\color{black} $s_{G,1}$ (the noisy version of the number of observations that are censored if $a = a_1$) is close to $n$}. Thus, $S_0(\boldsymbol{D}) \approx S_0(\boldsymbol{D}, a)$ implies that $S_L(\boldsymbol{D}) \approx S_L(\boldsymbol{D}, a) = S_0(\boldsymbol{D}, a) + \boldsymbol{\varepsilon}^{L}$ and
$
p_{\rm Dir}(\boldsymbol{x}  \mid  \boldsymbol{\alpha} = {\rm MLE}(S_L(\boldsymbol{D})-\boldsymbol{\varepsilon}^{L})) \approx
p_{\rm Dir}(\boldsymbol{x} \mid  \boldsymbol{\alpha} = {\rm MLE}(S_L(\boldsymbol{D},a)-\boldsymbol{\varepsilon}^{L})). 
$

Since the distribution of $\boldsymbol{\varepsilon}^{L}$ is known {\color{black} and $S_L(\boldsymbol{D},a)$ is observed, we know that the realization of $ \boldsymbol{\varepsilon}^{L}$ used in computing $S_L(\boldsymbol{D},a) $ satisfies $S_L(\boldsymbol{D},a) - \boldsymbol{\varepsilon}^{L} \in \text{Range}(S_0)$. This implies that when inferring $S_0(\boldsymbol{D}, a)$ from the distribution of $S_L(\boldsymbol{D},a) - \boldsymbol{\varepsilon}^{\ast L}$, it must be truncated to $\text{Range}(S_0)$. A further justification for the bootstrap algorithm can be found in Section~\ref{sec:boot}.} 


%

\begin{algorithm}
  \caption{DP parametric bootstrap to estimate $\boldsymbol{\alpha}$}
  \label{alg:MLE result}
  \begin{algorithmic}[0]
    \Procedure{DPBoots}{} {($S_{L}(\boldsymbol{D},a),\epsilon_1,a, n$)}
    \Repeat
        \begin{itemize}[itemindent=8mm]
            \item[i)] draw $\varepsilon^{\ast L}_j \overset{\mathrm{iid}}{\sim} {\rm Laplace}(0, - d \log(a)/(n \epsilon_1))$, $j=1,\, ... \, ,d$
        \item[ii)] compute $ \widetilde{S_0}(\boldsymbol{D},a) = (\tilde s_{0,1},\, ... \, ,\tilde s_{0,d}) = S_{L}(\boldsymbol{D},a) - (\varepsilon^{\ast L}_1,\dots,\varepsilon^{\ast L}_d)$
        \item[] if $\sum_{j=1}^d \exp(\tilde s_{0,j}) \leq 1$, proceed to iii); otherwise, return to i)
        \item[iii)] compute $\boldsymbol{\alpha}^* = {\rm MLE}(\widetilde{S_0}(\boldsymbol{D},a))$
        \item[iv)] generate simulated data set $\widetilde{\boldsymbol{D}}$ of size $n$ from ${\rm Dirichlet}( \boldsymbol{\alpha}^*)$
        \item[v)] obtain estimate $\widetilde{\boldsymbol{\alpha}} = {\rm MLE}(S_0(\widetilde{\boldsymbol{D}}))$
        \end{itemize}
      \Until reach a desired number of iterations\\
    \Return all $\widetilde{\boldsymbol{\alpha}}$'s obtained in all iterations
    \EndProcedure
  \end{algorithmic}
\end{algorithm}

\subsection{Bayesian methods} \label{sec:bayesian}

Under the Bayesian paradigm, inferences rely on the posterior distribution of  $\boldsymbol{\alpha}$. Due to privacy constraints, we assume that the only available information is $S_L(\boldsymbol{D},a)$, the DP version of the sufficient statistic.  

To use model (\ref{eq:BayesModel}), we need to treat either $\boldsymbol{D}$ or $S_0(\boldsymbol{D},a)$ as an unknown quantity and account for the noise added to it to define $S_L(\boldsymbol{D},a)$. As a result, an adequate inferential strategy must use the joint distribution of $(\boldsymbol{\alpha}, S_0(\boldsymbol{D},a))$ or $(\boldsymbol{\alpha}, \boldsymbol{D})$ conditional on $S_L(\boldsymbol{D},a)$, which is given by
\begin{eqnarray} \nonumber
& & \hspace{-20mm} \pi(\boldsymbol{\alpha}, \boldsymbol{D}|S_L(\boldsymbol{D},a) = \boldsymbol{s}_L)  \propto 
\\
\label{eq:DPposterior1a} & &
\prod_{j=1}^d p_{\rm Lap}\left(s_{L,j}\left|s_{0,j},-\frac{ d \log(a)}{n \epsilon_1}\right.\right)p(\boldsymbol{D}|\boldsymbol{\alpha}) \pi_0(\boldsymbol{\alpha}),
\end{eqnarray}
\begin{eqnarray} \nonumber
& & \hspace{-20mm} \pi(\boldsymbol{\alpha}, S_0(\boldsymbol{D},a)|S_L(\boldsymbol{D},a) = \boldsymbol{s}_L)  \propto 
\\\label{eq:DPposterior1b} & &
\prod_{j=1}^d p_{\rm Lap}\left(s_{L,j}\left|s_{0,j},-\frac{ d \log(a)}{n \epsilon_1}\right.\right) e^{(\boldsymbol{\alpha}-\boldsymbol{1})^T n S_0(\boldsymbol{D},a)-C(\boldsymbol{\alpha})} \pi_0(\boldsymbol{\alpha})
\end{eqnarray}
where $s_{0,j}$ and $s_{L,j}$ represent the $j$-th component of $S_0(\boldsymbol{D},a)$ and $\boldsymbol{s}_L$, respectively. Analysts will use $\pi(\boldsymbol{\alpha} |S_L(\boldsymbol{D},a) = \boldsymbol{s}_L)$, which is obtained by integrating out $\boldsymbol{D}$ in (\ref{eq:DPposterior1a}) or $S_0(\boldsymbol{D},a)$ in (\ref{eq:DPposterior1b}).  

We consider using a fraction of the data for prior elicitation to reduce the effects of prior choice on posterior inferences. Specifically, we partition $\boldsymbol{D}$ into two disjoint subsets $\boldsymbol{D}_1$ and $\boldsymbol{D}_2$, and use $S_{L}(\boldsymbol{D}_1,a)$ for prior elicitation and $S_{L}(\boldsymbol{D}_2,a)$ to define the likelihood. Under this alternative strategy, we use the joint distribution
\begin{eqnarray}\nonumber
& & \hspace{-12mm} \pi(\boldsymbol{\alpha}, S_0(\boldsymbol{D}_2,a)|S_L(\boldsymbol{D}_2,a) = \boldsymbol{s}_L^{(2)}, S_L(\boldsymbol{D}_1,a) = \boldsymbol{s}_L^{(1)})  \propto 
\\\label{eq:DPposterior2}
& & \hspace{-18mm} 
\prod_{j=1}^d p_{\rm Lap}\left(s_{L,j}^{(2)}\left|s_{0,j}^{(2)},\frac{- d \log(a)}{n_2 \epsilon_1}\right.\right) e^{(\boldsymbol{\alpha}-\boldsymbol{1})^T n_2 S_0(\boldsymbol{D}_2,a) -C(\boldsymbol{\alpha})} \pi_0(\boldsymbol{\alpha}|S_L(\boldsymbol{D}_1,a) = \boldsymbol{s}_L^{(1)})
\end{eqnarray}
where $s_{0,j}^{(2)}$, $s_{L,j}^{(1)}$, and $s_{L,j}^{(2)}$ represent the $j$-th component of $S_0(\boldsymbol{D}_2,a)$, $\boldsymbol{s}_L^{(1)}$, and $\boldsymbol{s}_L^{(2)}$, respectively. The prior $\pi_0$ in (\ref{eq:DPposterior2}) is informed by the subset $\boldsymbol{D}_1$ through the DP summary $S_L(\boldsymbol{D}_1,a)$. Inferences on $\boldsymbol{\alpha}$ can be made by integrating out $S_0(\boldsymbol{D}_2,a)$ in (\ref{eq:DPposterior2}). We discuss alternative approaches to specify $\pi_0(\boldsymbol{\alpha})$ and $\pi_0(\boldsymbol{\alpha}|S_L(\boldsymbol{D}_1,a) = \boldsymbol{s}_L^{(1)})$ in the next section.




\subsubsection{Prior distributions}
\label{sec:prior}
We consider five different approaches to define prior distributions on $\boldsymbol{\alpha}$. One of them is to assume that $\alpha_1, \, ... \, ,\alpha_d$ are independent and distributed according to a gamma distribution. Specifically, we assume that $\alpha_j \overset{\mathrm{ind.}}{\sim} {\rm Gamma}(v_j,w_j)$, $j \in \{1, 2, \, ... \, , d\}$, where $v_j$ and $w_j$ are shape and rate parameters, respectively. We define $\pi_0(\boldsymbol{\alpha}) = \prod_{j=1}^d p_{\rm Gam}(\alpha_j|v_j,w_j)$ and refer to it as prior p1. To define an uninformative prior, we set $v_j=1$ and $w_j=0.1$.

Given $S_L(\boldsymbol{D}_1,a) = \boldsymbol{s}_L^{(1)}$, we define $\pi_0(\boldsymbol{\alpha}|S_L(\boldsymbol{D}_1,a) = \boldsymbol{s}_L^{(1)})$ to be the distribution of $\boldsymbol{\alpha}$ induced by Algorithm \ref{alg:MLE result}. We refer to this prior as p2. We can draw values from p2 using the procedure ${\rm DPBoots}(S_{L}(\boldsymbol{D}_1,a),\epsilon_1,a$) described in Algorithm \ref{alg:MLE result}. Since prior p2 is unavailable in analytical form, we consider an additional prior p3 defined as  $\pi_0(\boldsymbol{\alpha}|S_L(\boldsymbol{D}_1,a) = \boldsymbol{s}_L^{(1)}) = \prod_{j=1}^d p_{\rm Gam}(\alpha_j|\hat{v}_j,\hat{w}_j)$ where $(\hat{v}_j,\hat{w}_j) = (\hat{v}_j (\boldsymbol{D}_1),\hat{w}_j(\boldsymbol{D}_1))$ are MLE estimates using a large random sample from p2. Under p3, we assume that $\alpha_1,\dots,\alpha_d$ are independent, which might not be necessarily the case under p2. To account for this potential dependence, we use copulas to define a joint distribution for $\boldsymbol{\alpha}$ while assuming that $\alpha_j \sim {\rm Gamma}(\hat{v}_j,\hat{w}_j)$. Specifically, we use a Gaussian copula \citep{sungur2000introduction} with a correlation matrix estimated using a large random sample from p2. We refer to this copula-based prior as p4. 

Finally, another choice of prior we consider, given $S_L(\boldsymbol{D}_1,a) = \boldsymbol{s}_L^{(1)}$, is to use model \ref{eq:DPposterior1a} and define 
$\pi_0(\boldsymbol{\alpha} \, | \, S_L(\boldsymbol{D}_1,a) = \boldsymbol{s}_L^{(1)}) =  \int \pi(\boldsymbol{\alpha}, \boldsymbol{D}_1 \, | \, S_L(\boldsymbol{D}_1,a) = \boldsymbol{s}_L^{(1)}) \, \mathrm{d} \boldsymbol{D}_1$. This prior is referred to as p5. We do not have an analytical expression for p5, so we are only able to draw from p5 using a sampling strategy, such as MCMC.

\subsubsection{Posterior inference based on MCMC}\label{sec:MCMCmethods}


First, we consider an MCMC algorithm based on the posterior distribution given by (\ref{eq:DPposterior1a}). In the algorithm, we must update both $\boldsymbol{\alpha}$ and $\boldsymbol{D}$. Unfortunately, it is not straightforward to sample from the conditional distributions $\boldsymbol{\alpha} \, | \, \boldsymbol{D},S_L(\boldsymbol{D},a)=\boldsymbol{s}_L$ and $\boldsymbol{D} \, | \, \boldsymbol{\alpha},S_L(\boldsymbol{D},a)=\boldsymbol{s}_L$, so we cannot implement a standard Gibbs sampler. We could use Metropolis-Hastings or slice sampling, but these samplers are computationally inefficient, particularly if the confidential data comprise hundreds or thousands of data points.
To overcome this computational issue, \cite{ju2022data} developed an MCMC algorithm that efficiently updates  $\boldsymbol{D}$ within each iteration using a one-variable-at-a-time Metropolis-Hastings algorithm. Its stationary distribution is equal to the posterior distribution $\pi(\boldsymbol{\alpha}|S_L(\boldsymbol{D},a)=\boldsymbol{s}_L)$.
We implement this algorithm with priors p1, p3, and p4 and refer to this approach as DPMCMC$\rm _{p1}$, DPMCMC$\rm _{p3}$, and DPMCMC$\rm _{p4}$, respectively.

The algorithm in \cite{ju2022data} can be computationally expensive for large $n$ because it needs to update $\boldsymbol{D}$. To reduce computation time, we borrow ideas based on data splitting that are commonly used to improve the scalability of Bayesian models \citep[see e.g.][]{
minsker;srivastava;lin;dunson;2014;paper, 
srivastava;li;dunson;2015}. The idea is to approximate the likelihood $p(\boldsymbol{D}|\boldsymbol{\alpha})$ by $\prod_{i=1}^b  [p_{\rm Dir}(\boldsymbol{x}_i|\boldsymbol{\alpha})]^{n/b}$, which is based on $b<n$ observations. To mimic a likelihood based on $n$ data points, each of the $b$ observations in the approximation is replicated $n/b$ times. Based on this approximation, and replacing $\boldsymbol{D}$ by $\boldsymbol{D}_1$, we use the model
\begin{eqnarray} \nonumber
& & \hspace{-20mm} \pi(\boldsymbol{\alpha}, [\boldsymbol{x}_1,\dots,\boldsymbol{x}_b]^T|S_L(\boldsymbol{D}_2,a) = \boldsymbol{s}_L^{(2)}, S_L(\boldsymbol{D}_1,a) = \boldsymbol{s}_L^{(1)})
\overset{\rm approx.}{\propto } \\
\label{eq:DPposterior1aprox} & &
\prod_{j=1}^d p_{\rm Lap}\left(s_{0,j}^{(2)}\left|s_{L,j}^{(2)},-\frac{ d \log(a)}{n_2 \epsilon_1}\right.\right) 
\prod_{i=1}^b  [p_{\rm Dir}(\boldsymbol{x}_i|\boldsymbol{\alpha})]^{n_2/b} \pi_0(\boldsymbol{\alpha}|S_L(\boldsymbol{D}_1,a) = \boldsymbol{s}_L^{(1)}),
\end{eqnarray}
We implement this modeling strategy using priors p1, p3 and p4, and refer to the resulting approaches as DPreMCMC$\rm _{p1}$, DPreMCMC$\rm _{p3}$ and DPreMCMC$\rm _{p4}$, respectively. The MCMC algorithm used to sample from these three approaches is a Metropolis-Hastings within Gibbs sampler where, for each $\boldsymbol{x}_i$, we use a one-variable-at-a-time Metropolis-Hastings algorithm with proposal distribution at time $t$ given by ${\rm Dirichlet}(\boldsymbol{\alpha}^{t-1})$ as in \cite{ju2022data} and, for $\boldsymbol{\alpha}$, we implement the slice sampler described in Figure 8 of \cite{neal2003slice}. 

We could also consider using MCMC techniques to sample from models (\ref{eq:DPposterior1b}) or  (\ref{eq:DPposterior2}). While we expect $S_0(\boldsymbol{D},a)$ to have a relatively small dimension relative to $\boldsymbol{D}$, it is not straightforward to characterize the distribution of $S_0(\boldsymbol{D},a)$, which is
a critical input for the implementation. 
Since this issue is hard to overcome, we decide to sample from models (\ref{eq:DPposterior1b}) and (\ref{eq:DPposterior2}) with ABC methods.

\subsubsection{Posterior inference with ABC}\label{sec:ABCmethods}

In the previous section, we argued that implementing MCMC algorithms to sample from models (\ref{eq:DPposterior1b}) or (\ref{eq:DPposterior2}) is unfeasible. To bypass this issue, we implement ABC approaches, which do not require evaluating either the likelihood or the prior \citep{tavare1997inferring}. 

In its most basic form, ABC is a rejection sampler that consists in drawing $\tilde{\boldsymbol{\alpha}} \sim \pi_0$, then $\tilde{\boldsymbol{x}}_i \, | \, \boldsymbol{\alpha} \overset{\mathrm{iid}}{\sim} {\rm Dirichlet} (\tilde{\boldsymbol{\alpha}})$, which we collect in a simulated data set denoted by $\tilde{\boldsymbol{D}}$, and finally drawing $\tilde{\varepsilon}^L_j \overset{\mathrm{iid}}{\sim} {\rm Laplace}(0, - d \log(a)/(n \epsilon_1))$, which we store in a vector $\tilde{\boldsymbol{\varepsilon}^L}$. After these simulations, we compute $S_L(\tilde{\boldsymbol{D}},a)=S_0(\tilde{\boldsymbol{D}},a)+\tilde{\boldsymbol{\varepsilon}^L}$, and we accept the draw $\tilde{\boldsymbol{\alpha}}$ if $\Vert S_L(\tilde{\boldsymbol{D}},a) - \boldsymbol{s}_L\Vert_2 < \delta$, with $\delta>0$. If the tolerance $\delta$ goes to zero, the accepted $\tilde{\boldsymbol{\alpha}}$ are exact draws from the posterior $\pi(\boldsymbol{\alpha}|S_L(\boldsymbol{D},a) = \boldsymbol{s}_L)$. Unfortunately, the acceptance rate of the algorithm is inversely related to $\delta$ and, in our case, it cannot be set to zero. To select $\delta$, we use the strategy proposed in \cite{pritchard1999population}, where $\delta$ is set to achieve an acceptance rate equal to a desired small value. We fix the acceptance rate at $0.1$.

We use ABC to sample under model (\ref{eq:DPposterior1b}) combined with the prior p1 and name this approach DPABC$\rm _{p1}$. Recall that p1 is an uninformative prior, which can negatively affect the accuracy of posterior inferences. Ideally, we prefer a prior that puts most of its probability mass in a region of the parameter space that is not too large. To make adequate comparisons, we also use ABC to sample under model (\ref{eq:DPposterior2}) with priors p2, p3, p4, and p5. We refer to these approaches as DPABC$\rm _{p2}$, DPABC$\rm _{p3}$, DPABC$\rm _{p4}$, and DPABC$\rm _{p5}$.

ABC has been previously applied to DP methods.  \cite{park2021abcdp} propose a DP algorithm that relies on sparse vector techniques. Two reasons dissuade us from using their methodology: i) $\epsilon$ depends on the number of accepted posterior draws, and ii) users cannot use additional DP-inferential methods without incurring in extra privacy loss.



\subsubsection{Posterior inference with asymptotic approximations} \label{sec:ASSYMPTOTICmethods}

Assuming the sample size is large enough, we can rely on asymptotic results to approximate the distribution of $S_0(\boldsymbol{D},a)$. In what follows, we assume that, as the sample size increases, smaller values of $a$ are used. To ensure that $S_L(\boldsymbol{D},a)$ is approximately equal to $S_0(\boldsymbol{D},a)$, we need to assume that as the sample size $n$ increases and $a$ decreases, 
$-d \log a/(n \epsilon_1)$ converges to zero. Since $S_0(\boldsymbol{D})$ is an average and the Dirichlet distribution is a member of the exponential family, we can use the central limit theorem to conclude that the asymptotic distribution of $S_0(\boldsymbol{D})$ is a multivariate normal distribution with mean $\boldsymbol{\mu}_{\boldsymbol{\alpha}}
    =(\Psi(\alpha_1),\, ... \, ,\Psi(\alpha_d))^T-\Psi(\sum_{j=1}^d \alpha_j) \boldsymbol{1}$ and covariance matrix $\boldsymbol{\Sigma}_{\boldsymbol{\alpha}}  =\mathrm{diag}(\Psi_1(\alpha_1),\, ... \, ,\Psi_1(\alpha_d)) -\Psi_1(\sum_{j=1}^d \alpha_j) \boldsymbol{1} \boldsymbol{1}^T$
where $\Psi$ is the digamma function, $\Psi_1$ is the trigamma function, and $\boldsymbol{1}$ is a $d$-dimensional vector of ones. {The approximation is full-dimension because $\boldsymbol{\Sigma}_{\boldsymbol{\alpha}}$ is equal to the Fisher information matrix of the Dirichlet distribution, which is invertible \citep{narayanan1991algorithm}}. We denote this asymptotic distribution as $p_{A}(S_0(\boldsymbol{D})|\boldsymbol{\mu}_{\boldsymbol{\alpha}},\boldsymbol{\Sigma}_{\boldsymbol{\alpha}})$. Under this approximation, we use the model
\begin{eqnarray} \label{eq:DPposterior1bApprox}
 \pi(\boldsymbol{\alpha}, S_0(\boldsymbol{D})|S_L(\boldsymbol{D},a) = \boldsymbol{s}_L)  \propto \prod_{j=1}^d p_{\rm Lap}\left(s_{0,j}\left|s_{L,j},-\frac{ d \log(a)}{n \epsilon_1}\right.\right) p_{A}(S_0(\boldsymbol{D})|\boldsymbol{\mu}_{\boldsymbol{\alpha}},\boldsymbol{\Sigma}_{\boldsymbol{\alpha}})
\end{eqnarray}
where $s_{0,j}$ represents the $j$-th component of $S_0(\boldsymbol{D})$.

Sampling directly from model (\ref{eq:DPposterior1bApprox}) is not straightforward. For that reason, we implement the Gibbs sampler algorithm proposed by \cite{bernstein2018differentially}, which draws directly from $S_0(\boldsymbol{D})|\boldsymbol{\alpha},S_L(\boldsymbol{D},a) = \boldsymbol{s}_L$. Updates from $\boldsymbol{\alpha}|S_0(\boldsymbol{D}), S_L(\boldsymbol{D},a) = \boldsymbol{s}_L$ are obtained using the Metropolis-Hastings algorithm discussed in Section \ref{sec:MCMCmethods}. To sample directly from $ S_0(\boldsymbol{D})|\boldsymbol{\alpha},S_L(\boldsymbol{D},a) = \boldsymbol{s}_L$, \cite{bernstein2018differentially} use the model augmentation approach developed by \cite{park2008bayesian} for the Bayesian LASSO.  \cite{bernstein2018differentially} also develop a strategy that accounts for censoring and, therefore, is even more ideal for our setup. Unfortunately, that strategy requires computing integrals that are not available in closed-form for the Dirichlet model. When attempting to approximate such integrals with numerical methods, we observe it is too computationally expensive to include this approximation within an MCMC scheme. 

We implement model (\ref{eq:DPposterior1bApprox}) combined with prior p1 and refer to it as DPapprox$\rm _{p1}$. In addition, we also consider an analogous asymptotic approximation for model (\ref{eq:DPposterior2}). This additional approximation uses priors p3 and p4, leading to two approaches that we refer to as DPapprox$\rm _{p3}$ and DPapprox$\rm _{p4}$.

\section{Simulation Study and Application}\label{sec:illustration}

In this section, we evaluate the performance of the methods in a simulation study and a real data set. We consider DPBoots and several Bayesian approaches. Based on the performance of the methods in the simulation study, we discard the methods that perform poorly. 

For convenience, we use the notation DPMCMC$\rm _{p1, p3, p4}$ to refer to DPMCMC combined with priors p1, p3, and p4. We use analogous notation for other computational strategies. The Bayesian approaches we consider are DPMCMC$\rm _{p1, p3, p4}$, DPreMCMC$\rm _{p1, p3, p4}$, DPABC$\rm _{p1, p2, p3, p4, p5}$, and DPapprox$\rm _{p1,p3,p4}$. Table \ref{tab:procedures} lists the prior distributions and targeted posteriors for each of the Bayesian approaches. 

\begin{table}[htbp]
  \centering
  \caption{Prior and posterior distributions for the DP Bayesian approaches}\label{tab:procedures}
\resizebox{.99\textwidth}{!}{
\begin{tabular}{c|c|c}
  \textbf{} & \textbf{Prior distribution} &\textbf{Posterior distribution}\\\hline
p1  &  Independent $\alpha_j \sim {\rm Gamma}(1,0.1)$ & $\pi\left(\boldsymbol{\alpha}|S_L(\boldsymbol{D},a) = \boldsymbol{s}_L\right)$\\\hline
p2  &  $\boldsymbol{\alpha} \overset{\rm distribution}{=} {\rm DPBoots}(S_{L}(\boldsymbol{D}_1,a),\epsilon_1,a)$ &
$\pi\left(\boldsymbol{\alpha}|S_L(\boldsymbol{D}_2,a) = \boldsymbol{s}_L^{(2)}, S_L(\boldsymbol{D}_1,a) = \boldsymbol{s}_L^{(1)}\right)$ \\\hline
p3  &  Independent $\alpha_j \sim {\rm Gamma}(\hat{v}_j (\boldsymbol{D}_1),\hat{w}_j(\boldsymbol{D}_1))$ &
$\pi\left(\boldsymbol{\alpha}|S_L(\boldsymbol{D}_2,a) = \boldsymbol{s}_L^{(2)}, S_L(\boldsymbol{D}_1,a) = \boldsymbol{s}_L^{(1)}\right)$ \\\hline
p4  & Gaussian copula and marginals ${\rm Gamma}(\hat{v}_j (\boldsymbol{D}_1),\hat{w}_j(\boldsymbol{D}_1))$ & $\pi\left(\boldsymbol{\alpha}|S_L(\boldsymbol{D}_2,a) = \boldsymbol{s}_L^{(2)}, S_L(\boldsymbol{D}_1,a) = \boldsymbol{s}_L^{(1)}\right)$ \\\hline
p5 & $\pi_0(\boldsymbol{\alpha}|S_L(\boldsymbol{D}_1,a) = \boldsymbol{s}_L^{(1)}) =  \int \pi(\boldsymbol{\alpha}, \boldsymbol{D}_1|S_L(\boldsymbol{D}_1,a) = \boldsymbol{s}_L^{(1)}) d\boldsymbol{D}_1$ & 
$\pi\left(\boldsymbol{\alpha}|S_L(\boldsymbol{D}_2,a) = \boldsymbol{s}_L^{(2)}, S_L(\boldsymbol{D}_1,a) = \boldsymbol{s}_L^{(1)}\right)$ 
\end{tabular}
}
\label{tab:tab1}
\end{table}
{\color{darkgreen}

}

\subsection{Simulation study}\label{subsec:numexp}

In the simulation study, we consider different values of $\boldsymbol{\alpha}$, sample sizes $n$, and privacy budgets $\epsilon$. For $\boldsymbol{\alpha}$, we consider $\boldsymbol{\alpha}^{\rm true}_1 = (3.3,4.4)^T$,
$\boldsymbol{\alpha}^{\rm true}_2 = (0.5,0.5,0.5)^T$,
$\boldsymbol{\alpha}^{\rm true}_3 = (2.2,3.3,4.4)^T$,
$\boldsymbol{\alpha}^{\rm true}_4 = (2,20,2)^T$, and 
$\boldsymbol{\alpha}^{\rm true}_5 = (2.2,3.3,4.4,5.5,6.6)^T$. For the sample sizes, we let $n\in\{1000,5000\}$. Finally, for the privacy budget we consider  $\epsilon\in\{0.25,0.5,1.5,10^{10}\}$ and $t_c = 0.01$. For the approaches based on data-splitting, we set $\epsilon_1 = 0.75 \epsilon$, $\epsilon_2 = 0.25\epsilon$,  $n_1 = 0.25n$, and $n_2 =0.75n$.  For each combination of $(\boldsymbol{\alpha},n,\epsilon)$, we simulate $50$ data sets and compute $S_{L}(\boldsymbol{D},a)$, $S_{L}(\boldsymbol{D}_1,a)$, and $S_{L}(\boldsymbol{D}_2,a)$. We select $a$ from a list of six candidates: $(0.1,10^{-2},10^{-3},10^{-4},10^{-5},10^{-6})$. In the case $\epsilon = 10^{10}$, there are virtually no privacy constraints, which allows us to evaluate directly the impact of the rescaling strategies (Section \ref{sec:MCMCmethods}), ABC (Section \ref{sec:ABCmethods}), and large sample approximations (Section \ref{sec:ASSYMPTOTICmethods}). As another benchmark, we consider an approach that targets the posterior distribution $\pi(\boldsymbol{\alpha}|S_0(\boldsymbol{D}))$ derived with no privacy constraints (i.e., model \ref{eq:BayesModel}) combined with prior p1, which we denote as MCMC$\rm _{p1}$. Similarly, the benchmark for DPBoots is parametric bootstrap with no privacy, which we refer to as Boots. 

{\color{black} For the bootstrap methods, we draw 1,000 values from the sampling distribution. To obtain 1,000 draws from the posterior distribution in each MCMC procedure, we run 3 chains, each with a length of 100,000 iterations and a burn-in of 20,000. For DPMCMC$\rm _{p1, p3, p4}$ and DPreMCMC$\rm _{p1, p3, p4}$, we found that the mixing improves as we update $\boldsymbol{\alpha}$ ``approximately'' from its full conditional distribution, which is not available in closed form. For that reason, instead of running one cycle of the slice sampler for $\boldsymbol{\alpha}$  within each MCMC iteration, we run 1,000 cycles of this slice sampler. The convergence of the MCMC chains was assessed using the Gelman-Rubin statistic. As expected, chain convergence depends on the sampler, prior distribution, and privacy budget, with DPapprox$\rm _{p1}$ exhibiting the slowest convergence. A detailed analysis regarding the convergence of the MCMC approaches is provided in the supplementary material (Section \ref{sec:supp_simulations}). We discard all instances for which the Gelman-Rubin statistic for $\boldsymbol{\alpha}$ is below 1.1, which is a common rule of thumb used to assess convergence.}

To assess the performance of the methods, we find the mean squared error (MSE) in estimating  $\boldsymbol{\alpha}$ and $E[\boldsymbol{x}|\boldsymbol{\alpha}] = \boldsymbol{\alpha}/(\sum_{j=1}^d \alpha_j)$. We also evaluate the coverage of the posterior predictive distributions for the Bayesian approaches.  


To evaluate the coverage of the posterior predictive distributions, we find approximate $95\%$ credible ellipsoids for each method and estimate the probability of falling within the ellipsoids with the true data-generating mechanisms. More precisely, we simulate data from the posterior predictive distribution and approximate the ellipsoids using the Mahalanobis distance centered at the estimated mean of the distribution. Then, we draw a large sample from the true data-generating mechanisms and calculate the fraction of data points that fall within the ellipsoids. If the computational strategies are accurate, the coverage of the ellipsoids should be close to $95\%$.

\begin{figure}[h]
    \centering
    \includegraphics[width=\textwidth, page = 1]{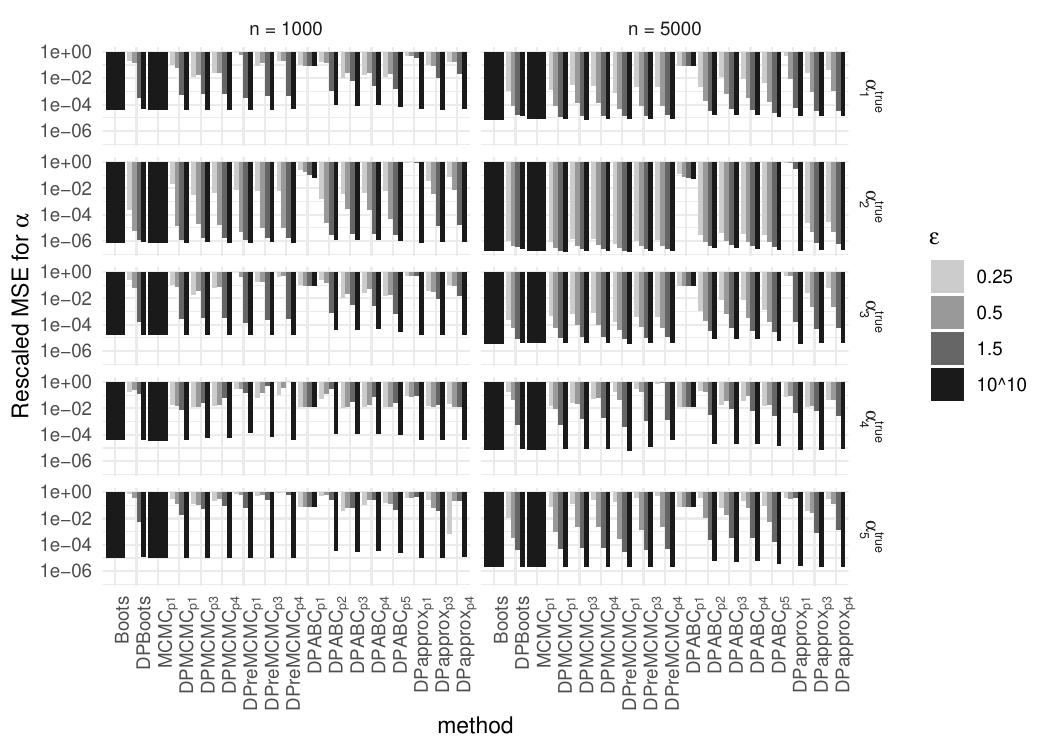}
    \caption{Rescaled median MSE for $\alpha$. The median is computed over all the MSE values obtained over the 50 simulated data sets and for each approach and combination of $(\boldsymbol{\alpha}^{\rm true},\epsilon,n)$. The non-private benchmarks MCMC and Boots are represented with the wider black bars.}
    \label{fig:MSE_all}
\end{figure}

Figure \ref{fig:MSE_all} displays the results we obtained for the MSE of $\boldsymbol{\alpha}$. To enhance the clarity of the results, we rescale the MSEs so that they are between 0 and 1 and present them in a logarithmic scale. Each bar represents the median MSE obtained across the 50 simulated data sets, with the bar starting at 1 and ending at the corresponding value, meaning that a larger bar corresponds to a smaller MSE. 

Our initial focus is on scenarios with virtually an unlimited privacy budget (i.e.,  $\epsilon = 10^{10}$). Among the methods considered, DPABC$\rm _{p1}$ has the weakest performance and we exclude it from further consideration. On the other hand, DPBoots, DPMCMC$\rm _{p1, p3, p4}$, DPreMCMC$\rm _{p1, p3, p4}$, and DPapprox$\rm _{p1, p3, p4}$ have MSEs that are similar to those of the benchmarks Boots and MCMC$\rm _{p1}$.  While DPABC$\rm _{p2, p3, p4, p5}$ have a slightly larger MSE, the performance is acceptable.

Now, we focus on the scenarios where $\epsilon \in \{0.25,0.5,1.5\}$. The MSEs are decreasing in $\epsilon$ and $n$ and increasing in $d$, which is to be expected. The methods DPapprox$\rm _{p1, p3, p4}$ produce slightly larger MSEs than the other methods and, for that reason, we discard them.

{\color{black} The results from DPABC$\rm _{p2, p3, p4, p5}$ show comparability, but we discard p3 because it does not incorporate prior dependence among $\boldsymbol{\alpha}$. We prefer DPABC$\rm _{p4, p5}$ over DPABC$\rm _{p2}$ because p4 allows more efficient sampling, and p5 is a prior defined in a principled manner using Bayes' theorem. In the case of DPreMCMC$\rm _{p1 p3, p4}$, performance is less satisfactory for $\boldsymbol{\alpha}^{\rm true}_4$ and $\boldsymbol{\alpha}^{\rm true}_5$ when $n = 1000$ or when $\epsilon = 0.25$. However, performance improves as $n$ increases to $5,000$, with similar performance to DPMCMC$\rm _{p1, p3, p4}$, particularly when $\epsilon = 1.5$. DPreMCMC$\rm _{p1}$ is chosen over the other two approaches because it is computationally faster than DPreMCMC$\rm _{p3, p4}$. 
Regarding DPMCMC$\rm _{p1, p3, p4}$, there is a slight advantage in using DPMCMC$\rm _{p1}$. DPBoots performs similarly to the best Bayesian approaches in terms of MSE.} All these observed discrepancies diminish when $n$ or $\epsilon$ increase. We now continue our analysis and comparisons using a representative for each of the Bayesian classes. Specifically, we choose DPMCMC$\rm _{p1}$, {\color{black} DPreMCMC$\rm _{p1}$, and DPABC$\rm _{p4,p5}$}.




Computation time is a key aspect to take into consideration, especially for the Bayesian approaches, as DPBoots is relatively fast. {\color{black} In the more challenging scenario with $n=5000$ and $\boldsymbol{\alpha}^{\rm true}_5$, 
DPMCMC$\rm _{p1}$ is the slowest approach among the selected ones.
Compared to DPMCMC$\rm _{p1}$, the approaches
DPreMCMC$\rm _{p1}$ ($b=5$) and DPABC$\rm _{p4,p5}$ are roughly 40\%, 99\%, and 50\% faster, respectively. If the sample size increases to 10000 or 100000, the most significant gain in computational speed is with DPreMCMC$\rm _{p1}$ (it is 60\% and 85\% faster, respectively).}

\begin{figure}[h]
    \centering
    \includegraphics[scale=0.9, page = 1, trim=0 0 30mm 0, clip]{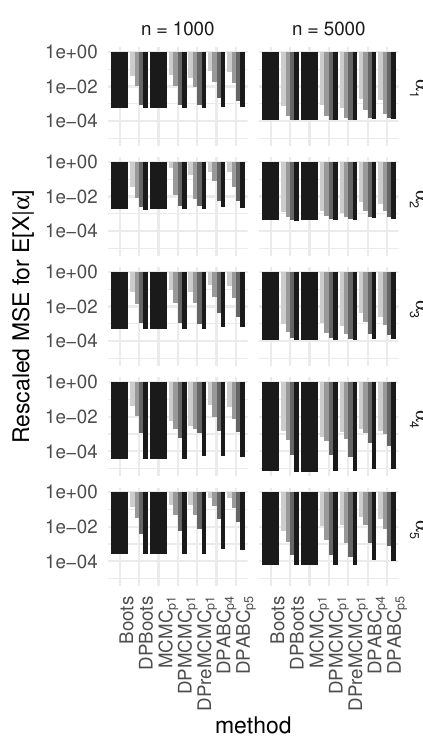} \hspace{5mm}
    \includegraphics[scale=0.9, page = 2]{fig_meanMSE_predCoverage.pdf}
    \caption{Left plot: Rescaled median MSE for $E[\boldsymbol{x}|\boldsymbol{\alpha}]$. The median is computed over all the MSE values obtained over the 50 simulated data sets and for each approach and combination of $(\boldsymbol{\alpha}^{\rm true},\epsilon,n)$. Right plot: Median posterior predictive coverage for the selected Bayesian approaches. The median is computed over all the coverage probabilities over the 50 simulated data sets and for each combination of $(\boldsymbol{\alpha}^{\rm true},\epsilon,n)$. In both plots, the non-private benchmarks MCMC and Boots are represented with the wider black bars.}
    \label{fig:ExpectedVal_Predictive}
\end{figure}

Figure \ref{fig:ExpectedVal_Predictive} shows the MSEs for $E[\boldsymbol{x}|\boldsymbol{\alpha}]$ and the posterior predictive coverages. The MSEs for $E[\boldsymbol{x}|\boldsymbol{\alpha}]$ are all similar, and they are generally better than the MSEs we found for $\boldsymbol{\alpha}$. 

The predictive coverage of the selected Bayesian methods {\color{black} is very similar, particularly when $n=5000$. Low coverage is particularly evident for DPreMCMC$_{p1}$ when $n=1000$ and $\epsilon \in \{0.25,0.5\}$. The coverage also suffers with almost all approaches when $n=1000$, $\boldsymbol{\alpha}^{\rm true}_5$, and $\epsilon \neq 10^{10}$.}

{\color{black}In addition to our main findings, we explore the use of the DP strategy for determining the threshold $a$, which is detailed in Section \ref{sec:DP_thershold}. We ran the same set of simulations with fixed $a$ and found that fixing $a$ can lead to numerical issues and higher MSEs for estimating $\boldsymbol{\alpha}$. Details for these additional simulations can be found in the supplementary material (Section~\ref{sec:supp_simulations}}).


We finish this section with some recommendations for users interested in implementing the methods. Users interested in frequentist approaches can use DPBoots, which performs fairly well. For those interested in Bayesian approaches, if $n$ is small, we recommend DPMCMC$\rm _{p1}$. Otherwise, if $n$ is large and $\epsilon \geq 1$, we recommend using DPreMCMC$\rm _{p1}$ or DPABC$\rm _{p4,p5}$. 
In general, users are advised to proceed with caution if $n$ is small and $\epsilon < 1$.

\color{black}

\subsection{Application to Daily Time Spent}

In this section, we apply our methods to a data set from the American Time Use Survey 2019 Microdata File\footnote{Publicly available at \href{https://www.bls.gov/tus/datafiles-2019.htm}{https://www.bls.gov/tus/datafiles-2019.htm}}, which we refer to as ATUS. 

ATUS is collected and housed by the U.S. Bureau of Labor Statistics, and it contains information on the daily time spent in of $18$ activities  (e.g., sex, personal care, household activities, and helping household members) during 2019. The data set contains $9435$ records released by the U.S. Bureau of Labor Statistics on July 22, 2021. 

We split up the data set between males and females and  define $\boldsymbol{x} = (x_1,x_2,x_3)$ as a compositional datum whose components are the fraction of time during the day spent on personal care ($x_1$), eating and drinking ($x_2$), and all other activities ($x_3$). After removing missing values and individuals that spend no time on personal care or eating and drinking, the sample size is $7125$ ($3359$ males and $3766$ females). 

Conceptually, we assume that ATUS is a confidential data set, and that an analyst wants to use DP to test for differences between females and males. Specifically, we assume the analyst wants to test if such differences are greater than $0.01$, that is, $H_{0,j}: |E[x_j|{\rm Male}] - E[x_j|{\rm Female}]| \leq 0.01$ versus $H_{1,j}: |E[x_j|{\rm Male}] - E[x_j|{\rm Female}]|  > 0.01$ for $j \in \{1,2,3\}$. 

We run the DP approaches DPBoots, DPMCMC$\rm _{p1}$, {\color{black} DPreMCMC$\rm _{p1}$ and DPABC$\rm _{p4,p5}$} a hundred times with $\epsilon=0.5$. We report our results in Table \ref{tab:tab7}. 
The approaches are independently run for males and females. By Proposition \ref{prop:parcomposition}, the privacy budget $\epsilon$ remains equal to $0.5$ after both analyses. To test these hypotheses, we use confidence intervals for DPBoots and posterior probabilities $\pi(H_{0,j}|S_L(\boldsymbol{D}_2,a) = \boldsymbol{s}_L^{(2)}, S_L(\boldsymbol{D}_1,a) = \boldsymbol{s}_L^{(1)})$ for the Bayesian approaches. We report the average (over the 100 runs) expected time estimate for each activity, the fraction of times that DPBoots rejects the null hypothesis at significance level $0.05$, and the fraction of times the Bayesian approaches have posterior probabilities of the null hypothesis below $0.5$.

The estimates for the expected values under DP are similar to the benchmarks. Regarding testing $H_{0,j}$, we find that, most of the time, the decision under DP and the benchmark is the same, with the exception of DPBoots and the activity related to eating and drinking. For this activity, while Boots rejects $H_{0,2}$, DPBoots fails to reject it. These types of discrepancies are common when testing hypotheses under DP because it injects additional uncertainty, which decreases the power of the tests. For Bayesian approaches, this phenomenon leads to posterior probabilities of $H_{0,j}$ shrinking to 0.5. In this application, we also experimented with different values of $\epsilon > 0.5$ and found that $\epsilon$ needs to be greater or equal to $1.25$ in order to make the decisions of Boots and DPBoots coincide. For all Bayesian approaches, the results for all hypotheses remain the same for  $\epsilon \geq 0.5$, and the conclusion is that there is evidence of gender-based differences in the time spent on personal care and other activities, while there is no evidence of differences when eating and drinking. We also observe that in DP Bayesian approaches, the posterior probabilities of $H_{0,j}$ approach the results with MCMC$\rm _{p1}$ as $\epsilon$ increases.


\begin{table}[t]
  \centering
  \caption{Average mean time estimates for males and females and fraction of times that the null hypothesis is rejected (over 100 repetitions). Prob column shows the average (over 100 repetitions) posterior probability of the null hypothesis.}
  \resizebox{17cm}{!}{
    \begin{tabular}{|c|c|r|c|c|r|c|c|r|c|c|}
    \hline
    \multirow{2}[4]{*}{Method} & \multirow{2}[4]{*}{Gender} & \multicolumn{3}{c|}{Personal Care} & \multicolumn{3}{c|}{Eating and drinking} & \multicolumn{3}{c|}{Other activities} \bigstrut\\
\cline{3-11}      &   & \multicolumn{1}{l|}{Mean} & \multicolumn{1}{l|}{Fraction} &   
\multicolumn{1}{l|}{Prob} & \multicolumn{1}{l|}{Mean} & \multicolumn{1}{l|}{Fraction} & 
\multicolumn{1}{l|}{Prob} & \multicolumn{1}{l|}{Mean} & \multicolumn{1}{l|}{Fraction} &  
\multicolumn{1}{l|}{Prob} \bigstrut\\
    \hline
    \multirow{2}[4]{*}{Boots} & Female & 0.411 & \multirow{2}[4]{*}{1} & \multirow{2}[4]{*}{} & 0.0507 & \multirow{2}[4]{*}{0} & \multirow{2}[4]{*}{} & 0.538 & \multirow{2}[4]{*}{1} & \multirow{2}[4]{*}{} \bigstrut\\
\cline{2-3}\cline{6-6}\cline{9-9}      & Male & 0.392 &   &   &  0.0508 &   &   & 0.557 &   &  \bigstrut\\
    \hline
    \multirow{2}[4]{*}{DPBoots} & Female & 0.411 & \multirow{2}[4]{*}{1} & \multirow{2}[4]{*}{} & 0.0513 & \multirow{2}[4]{*}{1} & \multirow{2}[4]{*}{} & 0.538 & \multirow{2}[4]{*}{1} & \multirow{2}[4]{*}{} \bigstrut\\
\cline{2-3}\cline{6-6}\cline{9-9}      & Male & 0.391 &   &   &  0.0518 &   &   & 0.557 &   &  \bigstrut\\
    \hline
    \multirow{2}[4]{*}{MCMC$\rm _{p1}$} & Female & 0.411 & \multirow{2}[4]{*}{1} & \multirow{2}[4]{*}{$10^{-5}$} &  0.0508 & \multirow{2}[4]{*}{0} & \multirow{2}[4]{*}{1} & 0.538 & \multirow{2}[4]{*}{1} & \multirow{2}[4]{*}{$10^{-5}$} \bigstrut\\
\cline{2-3}\cline{6-6}\cline{9-9}      & Male & 0.392 &   &   &  0.0510 &   &   &  0.557 &   &  \bigstrut\\
    \hline
    \multirow{2}[4]{*}{DPMCMC$\rm _{p1}$} & Female & 0.411 & \multirow{2}[4]{*}{1} & \multirow{2}[4]{*}{0.162} & 0.0510 & \multirow{2}[4]{*}{0} & \multirow{2}[4]{*}{0.879} & 0.538 & \multirow{2}[4]{*}{1} & \multirow{2}[4]{*}{0.199} \bigstrut\\
\cline{2-3}\cline{6-6}\cline{9-9}      & Male & 0.391 &   &   & 0.0516 &   &   & 0.557 &   &  \bigstrut\\
    \hline
    \multirow{2}[4]{*}{DPreMCMC$\rm _{p1}$} & Female & 0.411 & \multirow{2}[4]{*}{1} & \multirow{2}[4]{*}{0.099} & 0.0476 & \multirow{2}[4]{*}{0} & \multirow{2}[4]{*}{0.904} & 0.541 & \multirow{2}[4]{*}{1} & \multirow{2}[4]{*}{0.138} \bigstrut\\
\cline{2-3}\cline{6-6}\cline{9-9}      & Male & 0.391 &   &   & 0.0475 &   &   & 0.561 &   &  \bigstrut\\
    \hline
    \multirow{2}[4]{*}{DPABC$\rm _{p4}$} & Female & 0.410 & \multirow{2}[4]{*}{1} & \multirow{2}[4]{*}{0.240} & 0.0512 & \multirow{2}[4]{*}{0.05} & \multirow{2}[4]{*}{0.675} & 0.538 & \multirow{2}[4]{*}{1} & \multirow{2}[4]{*}{0.238} \bigstrut\\
\cline{2-3}\cline{6-6}\cline{9-9}      & Male & 0.391 &   &   & 0.0517 &   &   & 0.557 &   &  \bigstrut\\
    \hline
    \multirow{2}[4]{*}{DPABC$\rm _{p5}$} & Female & 0.410 & \multirow{2}[4]{*}{1} & \multirow{2}[4]{*}{0.241} & 0.0527 & \multirow{2}[4]{*}{0.03} & \multirow{2}[4]{*}{0.738} & 0.537 & \multirow{2}[4]{*}{1} & \multirow{2}[4]{*}{0.249} \bigstrut\\
\cline{2-3}\cline{6-6}\cline{9-9}      & Male & 0.391 &   &   & 0.0535 &   &   & 0.556 &   &  \bigstrut\\
    \hline
    \end{tabular}}%
  \label{tab:tab7}%
\end{table}%

For the Bayesian approaches, we also check the posterior  predictive distributions under DP for males and females. The results can be found in the Supplementary Material (Section~\ref{sec:postpred}). 

\section{Discussion}\label{sec:conclusions}

This article compared several approaches for analyzing  compositional data under DP constraints that are based on the Dirichlet distribution. 
For frequentist inference, we recommend DPBoots, which has performed well in our experiments. For Bayesian inference, we recommend implementing DPMCMC$\rm _{p1}$ for small sample sizes, and DPreMCMC$\rm _{p1}$ and DPABC$\rm _{p4,p5}$ when the sample size is moderate-to-large and $\epsilon \geq 1$.  

A limitation of our work is the potential inadequacy of the Dirichlet distribution for some compositional data sets. Future work could explore alternative models.  {\color{black} Additionally, further research is needed to establish a principled algorithm for choosing the censoring rate $t_c.$ This parameter should be small enough to avoid significant bias, yet not so small that it substantially increases the variance of the Laplace mechanism. For now, we recommend that analysts use $t_c = 0.01$, which has shown reasonable performance empirically.}

\bibliographystyle{asa}
\typeout{}
\bibliography{reference}

\begin{thebibliography}{27}
\newcommand{\enquote}[1]{``#1''}
\expandafter\ifx\csname natexlab\endcsname\relax\def\natexlab#1{#1}\fi

\bibitem[{Aitchison(1982)}]{aitchison1982statistical}
Aitchison, J. (1982), \enquote{The statistical analysis of compositional data,}
  \textit{Journal of the Royal Statistical Society: Series B (Methodological)},
  44, 139--160.

\bibitem[{Awan and Slavkovic(2020)}]{Awan_Slavkovic_2020}
Awan, J.~A. and Slavkovic, A. (2020), \enquote{Differentially private inference
  for binomial data,} \textit{Journal of Privacy and Confidentiality}, 10.

\bibitem[{Bacon-Shone(2011)}]{bacon2011short}
Bacon-Shone, J. (2011), \enquote{A short history of compositional data
  analysis,} \textit{Compositional data analysis: Theory and applications},
  3--11.

\bibitem[{Barrientos et~al.(2019)Barrientos, Reiter, Machanavajjhala, and
  Chen}]{barrientos2019differentially}
Barrientos, A.~F., Reiter, J.~P., Machanavajjhala, A., and Chen, Y. (2019),
  \enquote{Differentially private significance tests for regression
  coefficients,} \textit{Journal of Computational and Graphical Statistics},
  28, 440--453.

\bibitem[{Bernstein and Sheldon(2018)}]{bernstein2018differentially}
Bernstein, G. and Sheldon, D.~R. (2018), \enquote{Differentially private
  {B}ayesian inference for exponential families,} \textit{Advances in Neural
  Information Processing Systems}, 31.

\bibitem[{Bernstein and Sheldon(2019)}]{bernstein2019differentially}
--- (2019), \enquote{Differentially private {B}ayesian linear regression,}
  \textit{Advances in Neural Information Processing Systems}, 32.

\bibitem[{Dwork et~al.(2006)Dwork, McSherry, Nissim, and
  Smith}]{dwork2006calibrating}
Dwork, C., McSherry, F., Nissim, K., and Smith, A. (2006), \enquote{Calibrating
  noise to sensitivity in private data analysis,} in \textit{Theory of
  cryptography conference}, Springer, pp. 265--284.

\bibitem[{Efron(2012)}]{efron2012bayesian}
Efron, B. (2012), \enquote{Bayesian inference and the parametric bootstrap,}
  \textit{The annals of applied statistics}, 6, 1971.

\bibitem[{Ferrando et~al.(2022)Ferrando, Wang, and
  Sheldon}]{ferrando2022parametric}
Ferrando, C., Wang, S., and Sheldon, D. (2022), \enquote{Parametric bootstrap
  for differentially private confidence intervals,} in \textit{International
  Conference on Artificial Intelligence and Statistics}, PMLR, pp. 1598--1618.

\bibitem[{Ghosh et~al.(2012)Ghosh, Roughgarden, and
  Sundararajan}]{ghosh2012universally}
Ghosh, A., Roughgarden, T., and Sundararajan, M. (2012), \enquote{Universally
  utility-maximizing privacy mechanisms,} \textit{SIAM Journal on Computing},
  41, 1673--1693.

\bibitem[{Inusah and Kozubowski(2006)}]{inusah2006discrete}
Inusah, S. and Kozubowski, T.~J. (2006), \enquote{A discrete analogue of the
  Laplace distribution,} \textit{Journal of statistical planning and
  inference}, 136, 1090--1102.

\bibitem[{Ju et~al.(2022)Ju, Awan, Gong, and Rao}]{ju2022data}
Ju, N., Awan, J.~A., Gong, R., and Rao, V.~A. (2022), \enquote{Data
  augmentation MCMC for Bayesian inference from privatized data,}
  \textit{Advances in neural information processing systems}, 35, 12732--12743.

\bibitem[{Karwa and Vadhan(2017)}]{karwa2017finite}
Karwa, V. and Vadhan, S. (2017), \enquote{Finite sample differentially private
  confidence intervals,} \textit{arXiv preprint arXiv:1711.03908}.

\bibitem[{Kulkarni et~al.(2021)Kulkarni, J{\"a}lk{\"o}, Koskela, Kaski, and
  Honkela}]{kulkarni2021differentially}
Kulkarni, T., J{\"a}lk{\"o}, J., Koskela, A., Kaski, S., and Honkela, A.
  (2021), \enquote{Differentially private Bayesian inference for generalized
  linear models,} in \textit{International Conference on Machine Learning},
  PMLR, pp. 5838--5849.

\bibitem[{Minka(2000)}]{minka2000estimating}
Minka, T. (2000), \enquote{Estimating a Dirichlet distribution,} Tech. rep.,
  Massachusetts Institute of Technology.

\bibitem[{Minsker et~al.(2017)Minsker, Srivastava, Lin, and
  Dunson}]{minsker;srivastava;lin;dunson;2014;paper}
Minsker, S., Srivastava, S., Lin, L., and Dunson, D.~B. (2017), \enquote{Robust
  and scalable Bayes via a median of subset posterior measures,}
  \textit{Journal of Machine Learning Research}, 18, 4488--4527.

\bibitem[{Narayanan(1991)}]{narayanan1991algorithm}
Narayanan, A. (1991), \enquote{Algorithm AS 266: Maximum likelihood estimation
  of the parameters of the Dirichlet distribution,} \textit{Journal of the
  Royal Statistical Society. Series C (Applied Statistics)}, 365--374.

\bibitem[{Neal(2003)}]{neal2003slice}
Neal, R.~M. (2003), \enquote{Slice sampling,} \textit{The Annals of
  Statistics}, 31, 705--767.

\bibitem[{Ongaro and Migliorati(2013)}]{ongaro2013generalization}
Ongaro, A. and Migliorati, S. (2013), \enquote{A generalization of the
  Dirichlet distribution,} \textit{Journal of Multivariate Analysis}, 114,
  412--426.

\bibitem[{Park et~al.(2021)Park, Vinaroz, and Jitkrittum}]{park2021abcdp}
Park, M., Vinaroz, M., and Jitkrittum, W. (2021), \enquote{ABCDP: Approximate
  {B}ayesian computation with differential privacy,} \textit{Entropy}, 23, 961.

\bibitem[{Park and Casella(2008)}]{park2008bayesian}
Park, T. and Casella, G. (2008), \enquote{The Bayesian LASSO,} \textit{Journal
  of the American Statistical Association}, 103, 681--686.

\bibitem[{Pe{\~n}a and Barrientos(2021)}]{pena2021differentially}
Pe{\~n}a, V. and Barrientos, A.~F. (2021), \enquote{Differentially private
  methods for managing model uncertainty in linear regression models,}
  \textit{arXiv preprint arXiv:2109.03949}.

\bibitem[{Pritchard et~al.(1999)Pritchard, Seielstad, Perez-Lezaun, and
  Feldman}]{pritchard1999population}
Pritchard, J.~K., Seielstad, M.~T., Perez-Lezaun, A., and Feldman, M.~W.
  (1999), \enquote{Population growth of human Y chromosomes: a study of Y
  chromosome microsatellites.} \textit{Molecular biology and evolution}, 16,
  1791--1798.

\bibitem[{Skorski(2023)}]{skorski2023bernstein}
Skorski, M. (2023), \enquote{Bernstein-type bounds for beta distribution,}
  \textit{Modern Stochastics: Theory and Applications}, 10, 211--228.

\bibitem[{Srivastava et~al.(2018)Srivastava, Li, and
  Dunson}]{srivastava;li;dunson;2015}
Srivastava, S., Li, C., and Dunson, D.~B. (2018), \enquote{Scalable Bayes via
  barycenter in Wasserstein space,} \textit{Journal of Machine Learning
  Research}, 19, 312--346.

\bibitem[{Sungur(2000)}]{sungur2000introduction}
Sungur, E.~A. (2000), \enquote{An introduction to copulas,} \textit{Journal of
  the American Statistical Association}, 95, 334--334.

\bibitem[{Tavar{\'e} et~al.(1997)Tavar{\'e}, Balding, Griffiths, and
  Donnelly}]{tavare1997inferring}
Tavar{\'e}, S., Balding, D.~J., Griffiths, R.~C., and Donnelly, P. (1997),
  \enquote{Inferring coalescence times from DNA sequence data,}
  \textit{Genetics}, 145, 505--518.

\end{thebibliography}

\newpage


\makeatletter 
\renewcommand{\thefigure}{S\@arabic\c@figure}
\renewcommand{\thetable}{S\@arabic\c@table}
\renewcommand{\thesection}{S\@arabic\c@section}
\makeatother
\setcounter{figure}{0}
\setcounter{table}{0}
\setcounter{section}{0}

\begin{center}
{\Large \textbf{Supplementary Material: Differentially Private Inference for Compositional Data}}  \\
\end{center}

\noindent This document contains supplemental materials to accompany the main manuscript. The contents are:
\begin{itemize}
    \item In Section~\ref{sec:supp_cens}, we study the effects of censoring the sufficient statistic in our inferences. 
    \item In Section~\ref{sec:opta}, there is the proof of Theorem 1 in the main text, which establishes the convergence for the DP algorithm to select the censoring threshold $a$.
    \item In Section \ref{proof:theo1}, we provide the proof of Theorem 2 in the main text, which states the algorithm for releasing the DP statistic satisfies $(\epsilon_1+\epsilon_2)$-DP. 
    \item In Section~\ref{sec:boot}, we provide further justification for the bootstrapping scheme.
    \item  In Section \ref{sec:supp_simulations}, we include additional results for the simulation study.
    \item Finally, in Section \ref{sec:postpred}, we include two figures with the posterior predictive distributions estimated with the DP Bayesian methods for the ATUS application in the main text.
\end{itemize}

\section{Effects of censoring the sufficient statistic}
\label{sec:supp_cens}

In this section, we discuss the effects of censoring the sufficient statistic.  We do this by finding the expected proportion of censored entries and the bias induced by censoring. From our analysis, we conclude that inferences based on the censored data are close to what we would obtain without censoring if $a$ is small (relative to the expected value of $x_{ij}$) and $\sum_{k = 1}^d \alpha_k$ is not near zero. 

The probability that each entry $x_{ij}$ is censored is $
P( x_{ij} < a ) = P( \mathrm{Beta}(\alpha_j, \beta_j) < a) = I_{a}(\alpha_j, \beta_j),$  where $I_{a}(\cdot, \cdot)$ is the incomplete beta function and $ \beta_j = \sum_{k=1}^d \alpha_k - \alpha_j$. The higher $a$ is, the more likely an entry will be censored.

A quantity that summarizes the extent to which there are censored entries is the expected proportion of censored entries $x_{ij}$: 
$$
E\left[ \frac{1}{{n d}} \sum_{i = 1}^n \sum_{j = 1}^d \mathbbm{1}(x_{ij} < a) \right] = \frac{1}{n d} \sum_{i = 1}^n \sum_{j = 1}^d P(x_{ij} < a) =  \frac{1}{d} \sum_{j = 1}^d P(x_{ij} < a).
$$

\begin{figure}
\begin{subfigure}{.5\textwidth}
  \centering
  \includegraphics[width=\linewidth]{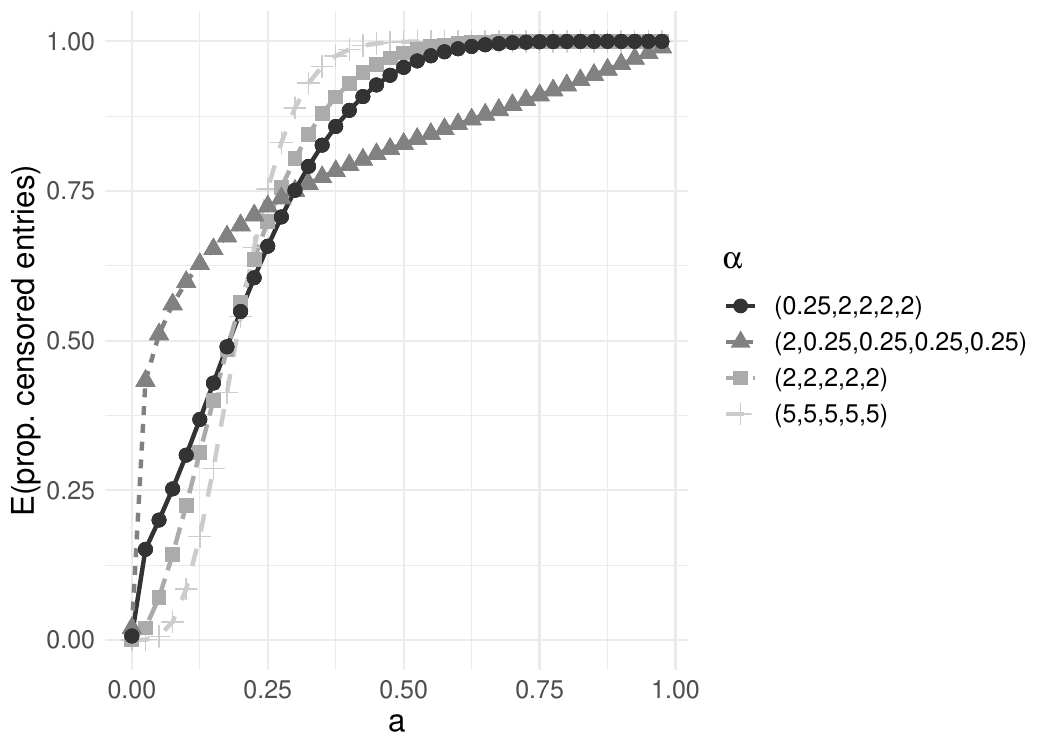}
  \caption{}
  \label{fig:cens}
\end{subfigure}%
\begin{subfigure}{.5\textwidth}
  \centering
  \includegraphics[width=\linewidth]{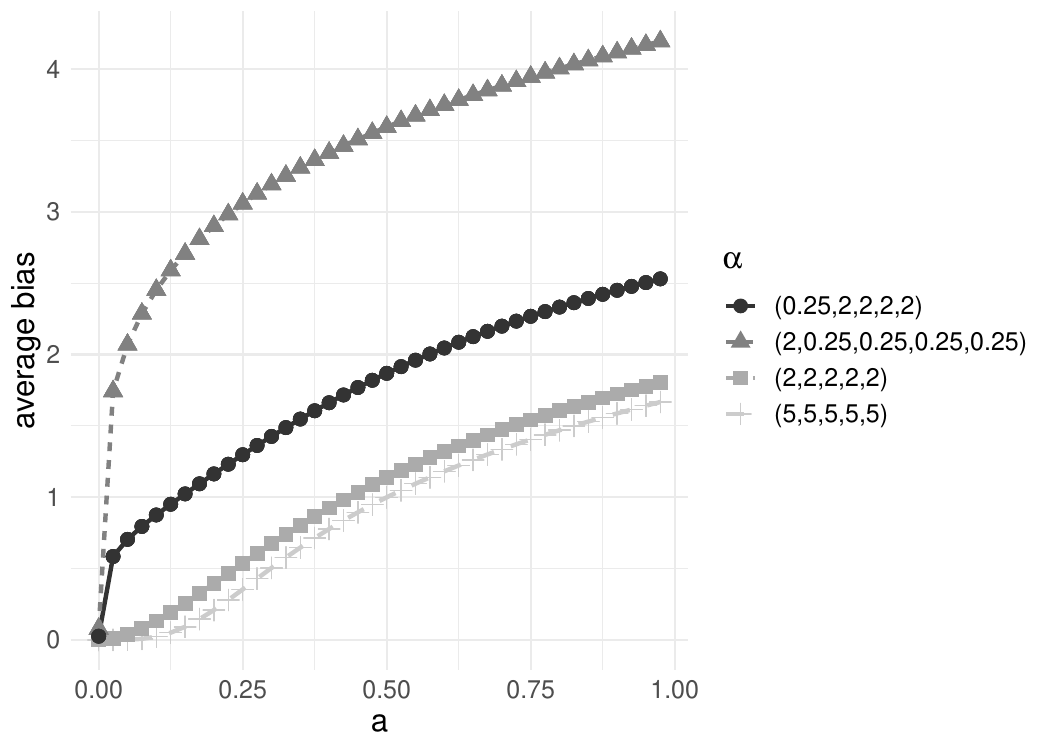}
  \caption{}
  \label{fig:bias}
\end{subfigure}
\caption{(a) Expected proportion of censored entries as a function of $\boldsymbol{\alpha}$ and $a$ (b)  Average bias of censored DP statistic as a function of $\boldsymbol{\alpha}$ and $a$.}
\label{fig:fig}
\end{figure}

Figure~\ref{fig:cens} shows the expected proportion of censored entries as a function of $a$ for different values of $\boldsymbol{\alpha}$. Unsurprisingly, the expected proportion increases in $a$. More interestingly, we observe that, as $\sum_{k=1} \alpha_k$ grows, the expected proportion is low for small values of $a$, but then increases more rapidly as $a$ increases. 

{\color{black} Another useful metric for quantifying the effects of censoring is the bias of the DP statistic. The DP and confidential censored statistics (without perturbation) share the same bias because the expectation of $\boldsymbol{\varepsilon}^L$ is zero, so $E[S_0(\boldsymbol{D},a)] = E[S_L(\boldsymbol{D},a)]$. }Figure~\ref{fig:bias} shows the average bias across the $d$ components of $\boldsymbol{x}_i$ as a function of $a$ for the same values of $\alpha$ we considered in Figure~\ref{fig:cens}. We observe that the bias increases in $a$ and decreases in $\sum_{k =1}^d \alpha_k$.

From an analytical point of view, we can bound the beta CDF with a Bernstein-type bound derived in \cite{skorski2023bernstein}. First of all, define
\begin{align*}
m_j = \frac{\alpha_j}{\sum_{k=1}^d \alpha_k}, \,  \,  v_j = \frac{\alpha_j \beta_j}{(\sum_{k=1}^d \alpha_k)^2 (\sum_{k=1}^d \alpha_k + 1)}, \, \, c_j = \frac{2 (\sum_{k=1}^d \alpha_k - 2\alpha_j)}{\sum_{k=1}^d \alpha_k (\sum_{k=1}^d \alpha_k + 2)}.
\end{align*}
Then, the probability that $x_{ij}$ is censored can be bounded as follows:
$$
P(x_{ij} < a) \le \begin{cases} e^{ \frac{-(m_j - a)^2}{2 [v_j + c_j(m_j - a)/3]}}, & \text{ if } \alpha_j \ge \sum_{k=1}^d \alpha_k/2 \\
e^{\frac{-(m_j - a)^2}{2 v_j}}, & \text{ if } \alpha_j < \sum_{k=1}^d \alpha_k/2. \end{cases}
$$
Most commonly, $\alpha_j < \sum_{k=1}^d \alpha_k/2$, unless the $j$-th component of $\alpha$ dominates over the others. Both cases are exponentially decreasing in $E(x_{ij}) - a$ provided that $E(x_{ij}) > a$, which in practice almost always holds because $a$ is chosen to be small.

Another metric we use to quantify the effects of censoring is the bias of the censored statistic. To obtain that quantity, we need to find the expected values of $S_0(\boldsymbol{D}, a)$ and $S_L(\boldsymbol{D}, a)$. The expectation of $\boldsymbol{\varepsilon}^L$ is zero, so $E[S_0(\boldsymbol{D},a)] = E[S_L(\boldsymbol{D},a)]$. The expected value of $\log \tilde{x}_{ij}$ is 
$$
E[\log \tilde{x}_{ij}] = E[\log x_{ij}] + P(x_{ij} < a) \log a - \mathcal{I}_a, \, \, \, \, \mathcal{I}_a = \int_{0}^a \log x_{ij} \, \mathrm{Beta}( x_{ij} \, | \, , \alpha_j, \beta_j) \, \mathrm{d} x_{ij}.
$$
The bias induced by censoring in the $j$-th component is 
$$
E[\log \tilde{x}_{ij}] - E[\log x_{ij}] = \int_{0}^a (\log a - \log x_{ij}) \, \mathrm{Beta}( x_{ij} \, | \, \alpha_j, \beta_j)  \, \mathrm{d} x_{ij} > 0.
$$
The bias is clearly increasing in $a$, as expected. In order to study the behavior of the bias analytically, we find an upper bound on the bias. Assuming $\alpha_j > 1$, we can use the bound $\log x_{ij} > 1-1/x_{ij}$ to do so:
\begin{align*}
E[\log \tilde{x}_{ij}] - E[\log x_{ij}] &\le  \int_{0}^a (\log a + 1/x_{ij} - 1) \, \mathrm{Beta}( x_{ij} \, | \, , \alpha_j, \beta_j)  \, \mathrm{d} x_{ij}\\
    &\le  \frac{B(\alpha_j-1, \beta_j)}{B(\alpha_j, \beta_j)} P(\mathrm{Beta}(\alpha_j-1, \beta_j) \le a),
 \end{align*}
 where $B( \cdot, \cdot)$ is the beta function. The ratio $B(\alpha_j-1, \beta_j)/B(\alpha_j, \beta_j)$ decreases to 1 as $\sum_{k = 1}^d \alpha_k$ increases, so the bound decreases as $\sum_{k = 1}^d \alpha_k$ increases. The beta tail probability can be bounded using the Bernstein-type bound we used in the previous section, which exponentially decreases in $E[x_{ij}] - a.$

\color{black}

\section{Proof of Theorem \ref{theo:opt_a}}
\label{sec:opta}
    Since $s_m \sim  {\rm Binomial}(n, p_m-p_{m-1})$, it is well-known that $s_m/n$ converges in probability to $p_m-p_{m-1}$ as $n$ goes to infinity. We now show that $s_{G,m}/n$ also converges to $p_m-p_{m-1}$ assuming $\epsilon = \Omega(n^{-\gamma})$. First, notice that
    \begin{eqnarray*}
        P\left[\frac{1}{n}\left|s_{G,m}-s_{m}\right|>t\right] & = & P\left[\frac{1}{n}\left|\max\left\{ s_{m}+\varepsilon_{m},0\right\} -s_{m}\right|>t\right]\\
         & = & P\left[\left|\max\left\{ s_{m}+\varepsilon_{m},0\right\} -s_{m}\right|^{2}>n^{2}t^{2}\right]\\
         & \leq & \frac{E\left[\left|\max\left\{ s_{m}+\varepsilon_{m},0\right\} -s_{m}\right|^{2}\right]}{n^{2}t^{2}}\\
         & \leq & \frac{E\left[\left|\max\left\{ s_{m}+\varepsilon_{m},0\right\} -s_{m}\right|^{2}\mathbb{I}\left(\varepsilon_{m}<-s_{m}\right)\right]}{n^{2}t^{2}}+\\
         &  & \hspace{5mm} \frac{E\left[\left|\max\left\{ s_{m}+\varepsilon_{m},0\right\} -s_{m}\right|^{2}\mathbb{I}\left(\varepsilon_{m}>-s_{m}\right)\right]}{n^{2}t^{2}}\\
         & \leq & \frac{E\left[\left|s_{m}\right|^{2}\mathbb{I}\left(\varepsilon_{m}<-s_{m}\right)\right]+E\left[\left|\varepsilon_{m}\right|^{2}\mathbb{I}\left(\varepsilon_{m}>-s_{m}\right)\right]}{n^{2}t^{2}}\\
         & \leq & \frac{E\left[s_{m}^{2}\mathbb{I}\left(\left|\varepsilon_{m}\right|>s_{m}\right)\mathbb{I}\left(s_{m}=0\right)\right]}{n^{2}t^{2}} + \\
         &  & \hspace{5mm}\frac{E\left[s_{m}^{2}\mathbb{I}\left(\left|\varepsilon_{m}\right|>s_{m}\right)\mathbb{I}\left(s_{m}>0\right)\right]+E\left[\varepsilon_{m}^{2}\right]}{n^{2}t^{2}}\\
         & \leq & \frac{E\left[E\left[s_{m}^{2}\mathbb{I}\left(\left|\varepsilon_{m}\right|>s_{m}\right)\mathbb{I}\left(s_{m}>0\right)\mid s_{m}\right]\right]+E\left[\varepsilon_{m}^{2}\right]}{n^{2}t^{2}}\\
         & \leq & \frac{E\left[s_{m}^{2}\left(E\left[\varepsilon_{m}^{2}\right]/s_{m}^{2}\mathbb{I}\left(s_{m}>0\right)\mid s_{m}\right)\right]+E\left[\varepsilon_{m}^{2}\right]}{n^{2}t^{2}}\\
         & = & \frac{2Var\left(\varepsilon_{m}\right)}{n^{2}t^{2}}
    \end{eqnarray*}
    where $\mathbb{I}(A)$ is an indicator function equal to 1 if the condition $A$ holds and 0 otherwise. Given $\epsilon_2 = \Omega(n^{-\gamma})$, then $Var\left(\varepsilon_{m}\right) = 2\exp(-\epsilon_2/2)/(1-\exp(-\epsilon_2/2))^{2}$ is $o(n^2)$ meaning $\left|s_{G,n}-s_{m}\right| = o_p(n)$, that is, $s_{G,m}/n$ also converges in probability to $p_m-p_{m-1}$. It follows that $n^{-1}\sum_{l=1}^{m} s_{G,l}$ and $n^{-1}\sum_{l=1}^{M+1} s_{G,l}$ converges in probability to $p_m$ and $1$, respectively. By Slutsky's Theorem, we then observe that  $\hat p_{G,m} = (\sum_{l=1}^{M+1} \min\{s_{G,m},0\})^{-1}(\sum_{l=1}^m \min\{s_{G,m},0\})$ converges in probability to $p_m$. 

    Let's consider the following two cases: i) $a_{\rm opt} = a_{\tilde m}$ for some $\tilde m \in \{1,\ldots,M\}$, that is, $p_m < t_c < p_{m+1}$ and ii) $a_{\rm opt} = 0$, that is, $t_c < p_{1}$. The proof is completed by noticing that, for cases i) and ii), 
    \begin{eqnarray*}
        P\left(a \neq a_{\rm opt}\right) & = &
        P\left(\{\hat{p}_{G,\tilde m} > t_c\} \cup
        \{\hat{p}_{G,\tilde{m}+1} < t_c \}\right) \\
        & \leq &
        P\left(\hat{p}_{G,\tilde m} - p_{\tilde m} > t_c-p_{\tilde m}\right) + 
        P\left(\hat{p}_{G,\tilde{m}+1} - p_{\tilde m+1}<
        t_c - p_{\tilde m+1}\right) \\
        & = &
        P\left(\hat{p}_{G,\tilde m} - p_{\tilde m} > t_c-p_{\tilde m}\right) + 
        P\left(-(\hat{p}_{G,\tilde{m}+1} - p_{\tilde m+1})>
        p_{\tilde m+1} - t_c\right) \\
        & \leq &
        P\left(|\hat{p}_{G,\tilde m} - p_{\tilde m}| > t_c-p_{\tilde m}\right) + 
        P\left(|\hat{p}_{G,\tilde{m}+1} - p_{\tilde m+1}|>
        p_{\tilde m+1} - t_c\right)
    \end{eqnarray*}
    and 
    \begin{eqnarray*}
        P\left(a \neq 0\right) & = &
        P\left(
        \hat{p}_{G,1} < t_c\right)\\
        & = &
        P\left(
        \hat{p}_{G,1} - p_{1} < t_c-p_{1}\right)\\
        & = &
        P\left(
        -(\hat{p}_{G,1} - p_{1}) > p_{1}-t_c\right)\\
        & \leq &
        P\left(
        |\hat{p}_{G,1} - p_{1}| > \frac{p_{1}-t_c}{2}\right)\\
    \end{eqnarray*}
    goes to 0 as $n$ increases, respectively, because  $\hat{p}_{G,m}$ converge in probability to $p_m$, $m \in \{1,\ldots,M\}$.
\color{black}

\section{Proof of Theorem 1}\label{proof:theo1}

Since $S_{G}(\boldsymbol{D})$ directly uses the Geometric mechanism, 
it satisfies $\epsilon_2$-DP. {\color{black} Notice that releasing \((\hat{p}_{G,1}, \ldots, \hat{p}_{G,M})\) and the threshold \(a\) does not incur any privacy loss as they are a function of only \(S_{G}(\boldsymbol{D})\) and because of the post-processing property of DP mechanisms (Proposition \ref{prop:postprocessing}).} If $n_1 = 0$ (i.e., without partitioning), $S_{L}(\boldsymbol{D},a)$ is $\epsilon_1$-DP by the Laplace mechanism. 
Thus, sequential composition (Proposition \ref{prop:seqcomposition}) ensures that releasing both $S_G(\boldsymbol{D})$ and $S_{L}(\boldsymbol{D},a)$ satisfies $(\epsilon_1+\epsilon_2)$-DP. If $n_1 > 0$ (i.e., with partitioning), releasing $S_{L}(\boldsymbol{D}_1,a)$ and $S_{L}(\boldsymbol{D}_2,a)$ is $\epsilon_1$-DP by the Laplace mechanism 
and parallel composition 
(Proposition \ref{prop:parcomposition}). 
Thus, sequential composition ensures that releasing $S_G(\boldsymbol{D})$, $S_{L}(\boldsymbol{D}_1,a)$, and $S_{L}(\boldsymbol{D}_2,a)$ satisfies $(\epsilon_1+\epsilon_2)$-DP.

\color{black}

\section{Data-generating mechanism for the bootstrap}
\label{sec:boot}

In this Section, we describe the data-generating mechanism that is involved in the bootstrap scheme. The data-generating mechanism is the convolution
 $$
 \int_{{\rm Range}({\rm Lap};\, S_L(\boldsymbol{D},a))} 
 p_{\rm Dir}(\boldsymbol{x} \mid \boldsymbol{\alpha} = {\rm MLE}(S_L(\boldsymbol{D},a)- \boldsymbol{\varepsilon}^L)) \, p_{\rm trunc-Lap}(\boldsymbol{\varepsilon}^{L} \mid  S_L(\boldsymbol{D},a), 0, m) \, \mathrm{d} \boldsymbol{\varepsilon}^{L},
 $$
 where $m = - d \log(a)/(n \epsilon_1)$,
 \begin{align*}
  {\rm Range}({\rm Lap};\, S_L(\boldsymbol{D},a)) & = \{\boldsymbol{\varepsilon}:\, S_L(\boldsymbol{D},a)- \boldsymbol{\varepsilon} \in {\rm Range}(S_0)\}, \\   
 {\rm Range}(S_0) & = \left\{(s_{0,1},\, ... \, ,s_{0,d}) \in (-\infty,0)^d \, : \, \sum_{j=1}^d \exp(s_{0,j}) \leq 1 \right\},
\end{align*}
 and
 $$
 p_{{\rm trunc-Lap}}(\boldsymbol{\varepsilon}^{L} \mid  S_L(\boldsymbol{D},a),0,m) = 
 \frac{\prod_{j=1}^{d} p_{\rm Lap}(\varepsilon^{L}_j|0,m)  \, \mathbbm{1} ( \boldsymbol{\varepsilon}^{L} \in {\rm Range}({\rm Lap};\, S_L(\boldsymbol{D},a)) )}{\int_{{\rm Range}({\rm Lap};\, S_L(\boldsymbol{D},a))} \prod_{j=1}^{d} p_{\rm Lap}(\varepsilon^{L}_j|0,m) \, \mathrm{d} \boldsymbol{\varepsilon}^{L}}.
$$
In the convolution, we use the truncated Laplace distribution $p_{{\rm Lap}}(\boldsymbol{\varepsilon}^{L}; S_L(\boldsymbol{D},a)|0,m)$. The reason is that $S_L(\boldsymbol{D},a)$ provides valuable information about the noise initially added to $S_0(\boldsymbol{D},a)$. This situation is analogous, for instance, to the case where the summary of interest is a count, and the observed noisy count is negative, which would inherently indicate that the added noise was not greater than zero.

The noise added to $S_0(\boldsymbol{D},a)$ due to privacy protection can be neutralized by subtracting a noise from $S_{L}(\boldsymbol{D},a)$ that follow the same distribution. Put another way, given the disclosed sufficient statistics, We can approximately extract $S_0(\boldsymbol{D},a)$ from $S_{L}(\boldsymbol{D},a)$ given by: $S_0(\boldsymbol{D},a) := S_{L}(\boldsymbol{D},a) - \boldsymbol{\varepsilon}^{L}$, where the entries of $\boldsymbol{\varepsilon}^{L}$ are independently sampled from ${\rm Laplace}(0, -d \log(a)/(n \epsilon_1))$. The bootstrap algorithm is needed to properly account for all sources of randomness, including the noise from the Laplace Mechanism and randomized within the MLE. When this is the case, there are two procedures to obtain $S_0(\boldsymbol{D},a)$ as the given observation in MLE. In each bootstrap replication, $S_0(\boldsymbol{D},a)$ can be obtained by easing up the impact of privacy protection from $S_{L}(\boldsymbol{D},a)$, and then it can be calculated using the data sampled from the estimated Dirichlet distribution. Now we introduce the estimation steps in our algorithm, DPBoots, summarized in Algorithm 3.

\section{Additional simulation results} \label{sec:supp_simulations}

\begin{figure}
 \centering
 \includegraphics[width=0.9\linewidth]{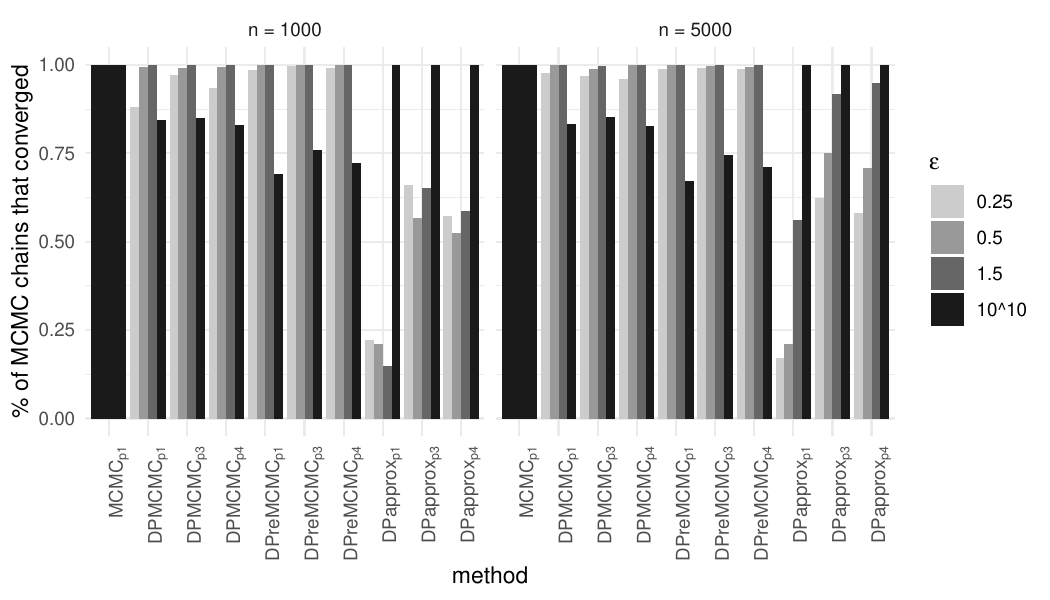}
 \caption{Fraction of times across simulated data sets and scenarios that the multivariate Gelman-Rubin statistic for $\boldsymbol{\alpha}$ is below 1.1.}
 \label{fig:convergence}
\end{figure}

To assess the convergence of the MCMC algorithms, we employed the multivariate Gelman-Rubin statistic, computed over the three generated MCMC chains for $\boldsymbol{\alpha}$. Each MCMC chain consisted of 100,000 iterations. Figure \ref{fig:convergence} displays the fraction of times across simulated data sets and scenarios that the statistic is below 1.1, a common rule of thumb used to claim convergence.

As expected, MCMC$_{\rm p1}$ converged in all instances. Both DPMCMC and DPreMCMC showed no convergence issues when $\epsilon$ was either 0.5 or 1.5. However, with smaller $\epsilon$ values, only a small fraction of chains required additional iterations for convergence. Conversely, when $\epsilon$ was excessively large, DPMCMC and DPreMCMC experienced slower convergence. We conjecture that this is due to the dramatically decreased probability of accepting candidates when updating the augmented data, as noted in \cite{ju2022data}. We also consider this is not a significant concern, as it is unrealistic for users to be allowed such an outrageously large privacy budget. If such a privacy budget were permitted, the privacy level would be negligible, and users could instead use MCMC${\rm p1}$. DPapprox convergence all the times when the privacy budget was excessively large ($\epsilon=10^{10}$). However, for other considered $\epsilon$ values, DPapprox frequently experiences slow convergence, except for DPapprox$_{\rm p3,p4}$ when $n=5000$.

Another aspect of interest is related to the use of the DP strategy for choosing $a$ outlined in Subsection \ref{sec:DP_thershold}. When the threshold is determined using this DP strategy, we refer to it as DP $a$. The idea is to study the effect of using a DP version of $a$  versus fixing it. More specifically, users might prefer to fix $a$ at a given value, instead of using the proposed DP strategy, thereby saving some of the privacy budget that can be used to query a less noisy version of $S_L(\boldsymbol{D},a)$.

Our goal is to compare the accuracy of the inference when using DP $a$ versus fixed $a$. In the simulations, DP $a$ uses 25\% of the privacy budget, with the remaining 75\% used to get $S_L(\boldsymbol{D},a)$. If the threshold $a$ is fixed, then the idea is to use the entire privacy budget to get $S_L(\boldsymbol{D},a)$. To assess the effect of using DP $a$ versus fixed $a$, we run the same simulations as when using DP $a$ of fixed $a$. Then, we compare the Mean Squared Error (MSE) for $\boldsymbol{\alpha}$ obtained for DP $a$ and fixed $a$. Recall that we select $a$ from a list of six candidates $\{10^{-6}, 10^{-5}, 10^{-4}, 10^{-3}, 10^{-2}, 0.1\}$. We decided to fix $a$ at two different values: $0.1$ and $10^{-6}$. The case $a = 10^{-6}$ represents a scenario where the user attempts to reduce potential bias while accepting a noisier version of $S_L(\boldsymbol{D},a)$.  The case $a = 0.1$ represents a scenario where the user is willing to accept potential bias while reducing the sensitivity of $S_L(\boldsymbol{D},a)$. We do not consider the scenario where the user has perfect information about where $a$ must be fixed. In such a scenario, we expect that the inferences should be more accurate compared to those based on DP $a$.

\begin{figure}
 \centering
 \includegraphics[width=0.9\linewidth]{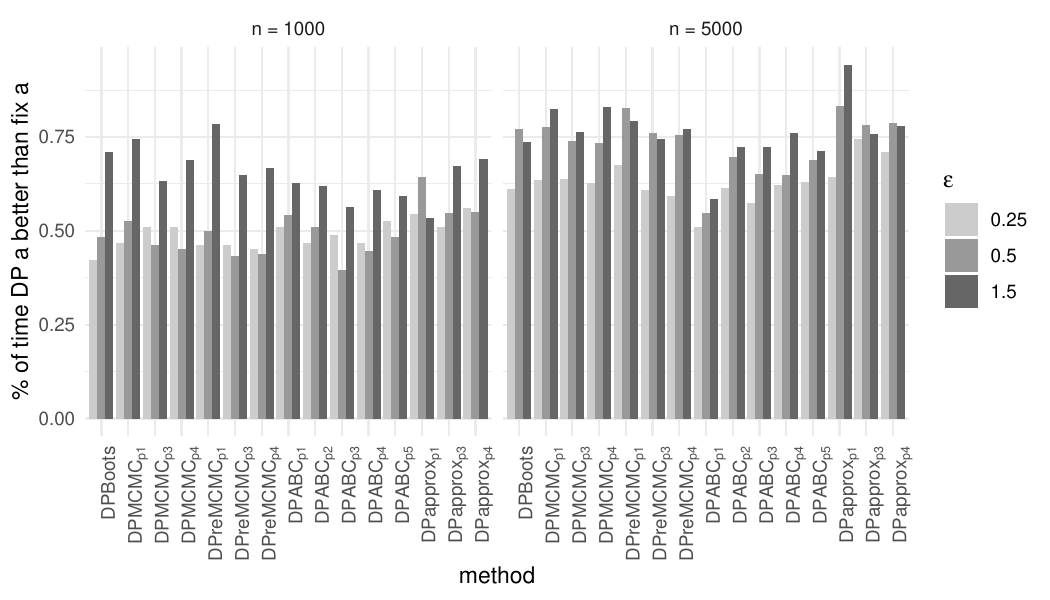}
 \caption{Fraction of times across simulated data sets and scenarios that the MSE for $\boldsymbol{\alpha}$ based a DP version of the threshold $a$ is smaller than that based on $a$ fixed at $10^{-6}$.}
 \label{fig:small_a}
\end{figure}

Figure \ref{fig:small_a} displays the fraction of times across simulated data sets and scenarios where the MSE for $\boldsymbol{\alpha}$ based on a DP version of the threshold $a$ is smaller, with $a$ fixed at $10^{-6}$. The figure presents results for $\epsilon \in \{0.25, 0.5, 1.5\}$, representing cases where splitting the privacy budget to obtain a DP version of $a$ is expected to impact inferences. For $n=5000$, the results favor using DP $a$, while for $n=1000$, a similar pattern is observed only when $\epsilon = 1.5$. When $n=1000$ and $\epsilon \in \{0.25, 0.5\}$, choosing between DP and fixed $a$ becomes less obvious. Sometimes using DP $a$ appears to be better, while other times fixed $a$ seems to be preferable. However, it is worth noting that the corresponding fractions of times associated with $n=1000$ and $\epsilon \in \{0.25, 0.5\}$ oscillate around 0.5.

For the case $a = 0.1$, we encounter numerical issues for $\boldsymbol{\alpha}^{\rm true}_4 = (2,20,2)^T$ and $\boldsymbol{\alpha}^{\rm true}_5 = (2.2,3.3,4.4,5.5,6.6)^T$. In the DPBoots algorithm, we must generate $(\tilde s_{0,1},, ... , ,\tilde s_{0,d}) = S_{L}(\boldsymbol{D},a) - (\varepsilon^{L}_1,\dots,\varepsilon^{L}_d)$ subject to $\sum_{j=1}^d \exp(\tilde s_{0,j}) \leq 1$, where $\varepsilon^L_j \overset{\mathrm{iid}}{\sim} {\rm Laplace}(0, -d \log(a)/(n \epsilon_1))$. To generate this random statistic, which is required to obtain a bootstrap draw of $\boldsymbol{\alpha}$, we employ a rejection sampler. The numerical issue arises when the acceptance probability in this rejection sampler is extremely small, which is expected to happen if $S_{L}(\boldsymbol{D},a)$ is far (in terms of the variance of $\varepsilon^L_j$) from ${\rm Range}(S_0)$. We observe that after several days of running this rejection sampling with $\boldsymbol{\alpha}^{\rm true}_4$ and $\boldsymbol{\alpha}^{\rm true}_5$, most of the simulations were not able to complete the desired bootstrap sample size. This is unappealing and indicates that users must act with caution if they want to fix $a$ at a ``large'' value that might lead to a large number of observations being censored, as was the case for $\boldsymbol{\alpha}^{\rm true}_4$ and $\boldsymbol{\alpha}^{\rm true}_5$, potentially resulting in numerical issues.

\color{black}
\section{Posterior predictive distributions for ATUS data} \label{sec:postpred}

In this section, we include the average posterior predictive distributions estimated with the DP Bayesian methods (averaged over the 100 runs).

\begin{figure}
 \centering
 \includegraphics[width=0.9\linewidth]{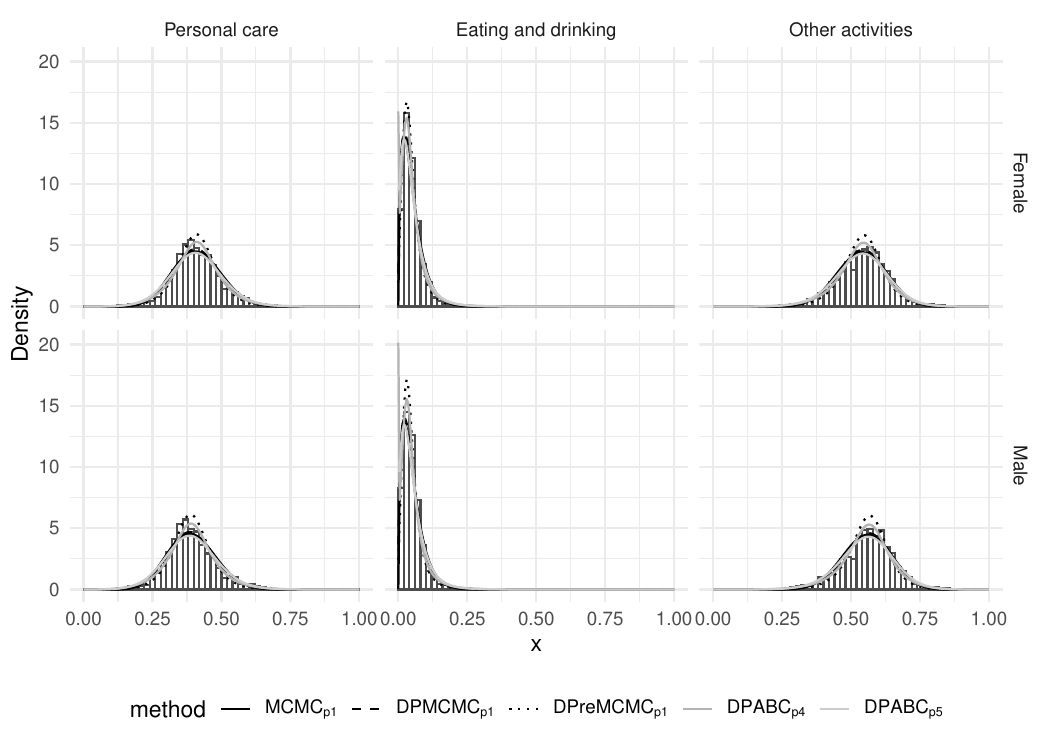}
 \caption{Posterior predictive distributions using DPreMCMC$\rm _{p4}$ with $\epsilon = 0.5$ and MCMCp1 (benchmark) for males and females and each activity. Histograms represent the observed data points.}
 \label{fig:fig3}
\end{figure}

\begin{figure}
 \centering
 \includegraphics[width=0.9\linewidth]{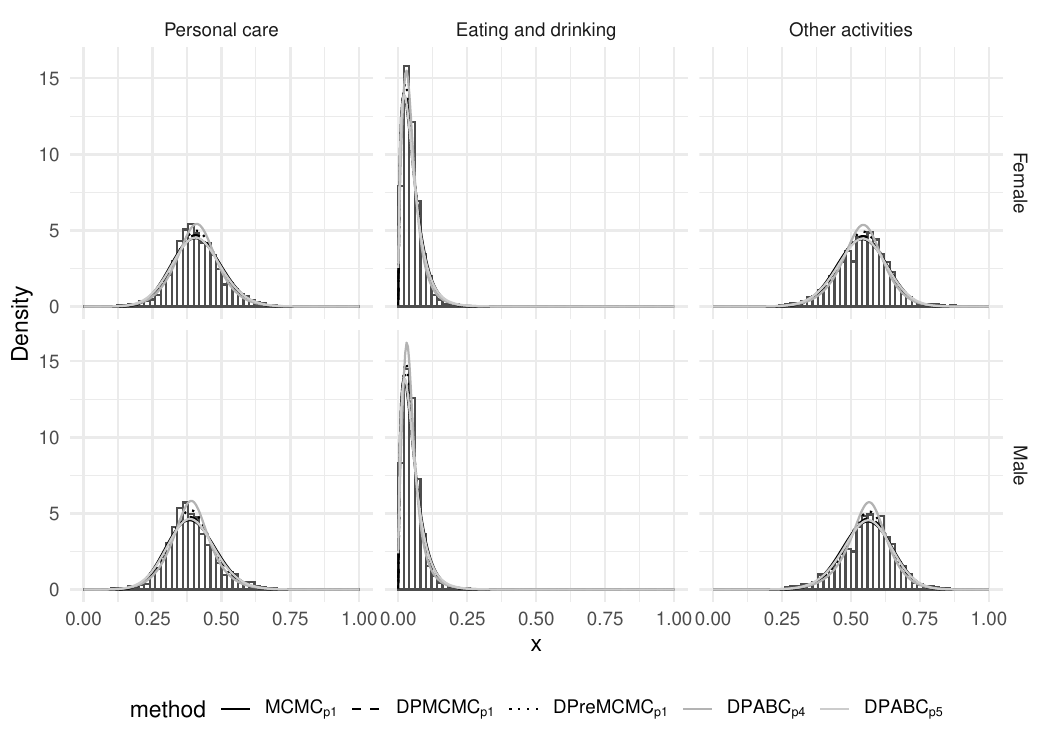}
 \caption{Posterior predictive distributions using DPreMCMC$\rm _{p4}$ (with $\epsilon = 1$) and MCMCp1 (benchmark) for males and females and each activity. Histograms represent the observed data points.}
 \label{fig:fig4}
\end{figure}

{\color{black} Figure \ref{fig:fig3} and \ref{fig:fig4} display the average of the estimated predictive distributions and histograms of the observed data for $\epsilon$ equal to 0.5 and 1, respectively. The posterior predictive densities are similar to the observed data in both Figures, and the density estimates under DP are similar to those obtained through MCMC$\rm _{p1}$. There is a small discrepancy in the density function for the activity ``eating and drinking'' when using DPABC$\rm _{p4,p5}$ with  $\epsilon = 0.5$. Noticeably, the value of this density function when a person spends no time eating and drinking is positive. These behaviors are commonly observed when estimating probability density functions in bounded intervals, particularly near the edges. If the goal is to estimate moments or quantiles, DPABC$\rm _{p4,p5}$ produces results similar to those using MCMC$\rm _{p1}$. These similarities are no longer maintained when estimating functionals such as the mode.}

\end{document}